\documentclass[conference,compsoc]{IEEEtran}
\usepackage[all]{pdfprivacy}

\ifCLASSOPTIONcompsoc
  \usepackage[nocompress]{cite}
\else
  \usepackage{cite}
\fi

\usepackage{graphicx}             % \includegraphics; detects pdfLaTeX itself, so no [pdftex]
\usepackage{amssymb}              % \mathbb{E}, \mathbb{I}
\usepackage{amsmath}              % \text and equations
\usepackage{xspace}               % spacing after \ourMethod
\usepackage{booktabs}             % \toprule, \midrule, \bottomrule
\usepackage{microtype}            % line breaking; removing it reflows the whole paper
\usepackage{tcolorbox}
\usepackage[table]{xcolor}
\usepackage{array}
\usepackage{multirow}
\usepackage{algorithm}
\usepackage{algpseudocode}
\usepackage{enumitem}
\usepackage{paralist}             % compactitem in the introduction
\usepackage{xurl}                 % URLs may break after any character
\usepackage[hidelinks]{hyperref}  % \url and clickable references; load last

\usepackage{todonotes}

\newcommand{\ourMethodNoSpace}{MASCing}
\newcommand{\ourMethod}{\ourMethodNoSpace\xspace}

\makeatletter
\newcommand{\linebreakand}{%
  \end{@IEEEauthorhalign}
  \hfill\mbox{}\par
  \mbox{}\hfill\begin{@IEEEauthorhalign}
}
\makeatother

\begin{document}

\title{MASCing: Configurable Mixture-of-Experts Behavior via Activation Steering Masks}
% MASCing: Configurable Safety Behaviors in Mixture-of-Experts Models via Activation Steering Masks

% \author{
%     \IEEEauthorblockN{Anonymous Author(s)}
% }

\author{ \IEEEauthorblockN{Jona te Lintelo} \IEEEauthorblockA{\textit{Radboud University}} \and \IEEEauthorblockN{Lichao Wu} \IEEEauthorblockA{\textit{University of Bristol}} \and \IEEEauthorblockN{Marina Kr\v{c}ek} \IEEEauthorblockA{\textit{Radboud University}} \linebreakand \IEEEauthorblockN{Sengim Karayal\c{c}in} \IEEEauthorblockA{\textit{Leiden University}} \and \IEEEauthorblockN{Stjepan Picek} \IEEEauthorblockA{ \textit{Faculty of Electrical Engineering and Computing,}\\ \textit{University of Zagreb \&}\\ \textit{Radboud University} } }

\maketitle

\begin{abstract}
Large language models (LLMs) operate under safety requirements that vary across applications and evolve with emerging threats and deployment policies. Conventional training-based interventions make adapting model behavior computationally expensive. Mixture-of-Experts (MoE) models introduce a further control challenge because each input activates only a subset of experts, coupling safety behavior to discrete routing decisions. Rapid adaption of MoE LLMs therefore requires a lightweight mechanism for identifying and controlling the routing decisions that produce the target behavior. We present MASCing (MoE Activation Steering Configuration), a framework that enables flexible configuration of safety-related MoE behavior without training the underlying model. MASCing uses an LSTM surrogate model to capture cross-layer dependencies and map routing logits to downstream behaviors. Guided by this surrogate, it identifies safety-relevant experts and optimizes a sparse steering mask to bias expert selection. This enables targeted enhancement or suppression of specific behaviors while preserving model utility. We evaluate MASCing on three contrasting safety objectives and observe consistent gains across seven MoE models. For multi-turn jailbreak defense, it improves the average defense success rate from 52.5\% to 83.9\%. For adult-content generation, MASCing enables models to comply with such requests that would otherwise be refused, increasing the average generation success rate from 52.6\% to 82.0\%. For political neutrality, MASCing makes models decline to state personal opinions on political questions, which they otherwise often give, raising the share of such questions on which they take no position from 31.8\% to 75.4\% while other questions remain answered. These results establish MASCing as a practical, lightweight, and flexible framework for task-specific safety configuration in MoE models.
\end{abstract}

\section{Introduction}
\label{sec:introduction}

Large Language Models (LLMs) have demonstrated impressive capabilities across natural language understanding, code generation, and logical reasoning~\cite{chen2021evaluatinglargelanguagemodels,brown2020fewshotlearners,kojima2022oneshotreasoners,wei2022emergent}. However, scaling dense LLMs is costly, as every parameter is activated for every input token, and larger models directly increase computational, memory, and infrastructure demands~\cite{hu2022lora,lialin2024scalingscaleupguide,patterson2022carbon}. Mixture-of-Experts (MoE) architectures address this bottleneck through conditional computation. An MoE model contains multiple expert networks and routes each token to only a small subset of them~\cite{shazeer2017outrageouslylargeneuralnetworks}. This allows the model to expand total parameter capacity while keeping the number of active parameters per forward pass relatively small. As a result, MoE models can achieve large-scale capacity at substantially lower inference cost than similarly sized dense models~\cite{shazeer2017outrageouslylargeneuralnetworks,fedus2022switchtransformers}. This efficiency has made MoE architectures increasingly prominent in modern LLM development, including models from Microsoft~\cite{abdin2024phi3moe}, OpenAI~\cite{openai2025gptossmoe}, DeepSeek~\cite{dai2024deepseekmoe}, Alibaba~\cite{yang2025qwen3moe}, and Mistral~\cite{jiang2024mixtralmoe}.

Although the MoE architecture offers significant cost-performance benefits, its unique structure also introduces new safety concerns. Beyond the jailbreak risks shared with dense LLMs~\cite{wei2023jailbroken,niu2024jailbreaking,shen2024anything}, the sparse expert-routing mechanism of MoE models can itself become an attack surface where adversaries can manipulate or suppress safety-relevant experts to bypass safety alignment and induce harmful outputs that would otherwise be refused~\cite{lai2025safex,wu2025gatebreaker,lintelo2026largelanguagelobotomy}. Unfortunately, most approaches for adapting or hardening a model's safety behavior are performed via training-based interventions, such as fine-tuning, alignment tuning, or retraining~\cite{lai2025safex}. These methods are expensive for large models and have been shown to be less effective for MoE models, where even benign fine-tuning can shift harmful inputs away from safety-critical experts~\cite{kim2026safemoe}. Moreover, safety requirements are dynamic: policies evolve, deployment contexts differ, and emerging misuse patterns may require quick adaptation. A defense that requires modifying or retraining the full model is thus poorly suited for fast and flexible safety control. As such, the AI safety community lacks a mechanism to rapidly configure the safety behavior of MoE LLMs across diverse and changing deployment contexts. 

In this paper, we introduce \ourMethod (MoE Activation Steering Configuration), a lightweight framework for configuring safety-related behavior in MoE models. \ourMethod is built around a three-phase pipeline: (i) Sequential Modeling of Behavior, (ii) Steering Mask Creation, and (iii) Steering Mask Application. First, to intervene on MoE behavior without retraining the model itself, we need a tractable representation of how safety-relevant behavior emerges through routing decisions. \ourMethod trains a long short-term memory (LSTM)~\cite{hochreiter1997lstm} surrogate on continuous, unnormalized routing logits. This preserves fine-grained routing information and allows the surrogate to capture both temporal patterns across token sequences and cross-layer dependencies. Next, to identify which parts of the routing mechanism drive the desired safety outcomes, \ourMethod uses the differentiable surrogate to search for the specific experts that must be enhanced or suppressed. We then optimize a continuous steering matrix and keep it sparse, so that the resulting interventions are limited to the few experts that drive the target behavior and the routing of all other experts stays unchanged. Finally, to configure model behavior at deployment time, \ourMethod converts these discovered interventions into a static steering mask and applies it directly to the routing logits. The mask shifts the pre-routing logits and thereby biases the top-$k$ selection toward the target behavior, while the model's behavior outside the target domain, its general capabilities, and the coherence of its responses stay largely intact. In this way, \ourMethod provides a practical mechanism for targeted configuration of safety-related behavior without modifying model weights or incurring the cost of retraining.

To demonstrate how flexibly \ourMethod can configure LLM safety behavior, we evaluate it on three contrasting objectives: enforcing security boundaries, selectively relaxing them, and restricting a domain that the model otherwise engages with. First, to validate its ability to enforce security boundaries, we address multi-turn jailbreak defense. This setting targets a practical vulnerability in conversational systems, where adversarial intent is gradually concealed over extended interactions. In this defensive scenario, \ourMethod improves the average defense success rate from 52.5\% to 83.9\%, reaching a maximum success rate of 89.2\%, while the share of benign prompts it refuses rises by at most 4.3\%. Against defense-aware adaptive attackers, \ourMethod raises the defense success rate from 30.7\% to 74.3\%. Second, to demonstrate controlled relaxation of security boundaries, we apply \ourMethod to enable adult-content generation. This reflects emerging policy trends in which platforms differentiate content access by user group or application context~\cite{openai2025adult}. Rather than broadly weakening safeguards, this scenario tests whether safety constraints can be relaxed in a targeted, policy-aligned manner. \ourMethod induces compliance for prompts the base model would typically refuse, increasing the average generation success rate from 52.6\% to 82.0\%, reaching up to 93.0\%, while the share of unrelated harmful requests answered with harmful content changes by at most 3.4\%. Third, to restrict a domain that a deployer wants to exclude, we apply \ourMethod to make models decline to state personal opinions on political questions, as deployers of assistants for a broad public may require for political neutrality. Unlike a refusal of harmful content, the steered model is expected to stay helpful by declining to take a side and offer neutral information. \ourMethod raises the share of political questions on which the models take no position from 31.8\% to 75.4\%, while the share of other questions that they decline changes by at most 4.5\%. Importantly, these adjustments incur minimal computational overhead. At inference time, the method only adds a steering mask to the routing logits, and the LSTM surrogate used for safety-relevant expert identification can be trained within five minutes on a single NVIDIA H100 GPU. Our contributions include:
\begin{compactitem}
    \item We propose \ourMethodNoSpace, a lightweight framework that configures safety-related behavior in MoE models using sparse steering masks on the routing logits, without fine-tuning the model and while preserving model utility.
    \item We introduce a method for analyzing MoE behavior through continuous routing logits rather than discrete expert selections, which enables a differentiable LSTM surrogate that identifies the experts associated with a target behavior by leveraging sequential information.
    \item We show across seven MoE models that \ourMethod configures safety-related behavior in several directions. It strengthens multi-turn jailbreak defense without introducing high over-refusal, selectively relaxes refusal for adult content while preserving refusal of unrelated harmful requests, and restricts personal opinions on political questions while other questions remain answered.
\end{compactitem}
% \textbf{Paper Organization.} The remainder of this paper is organized as follows: Section~\ref{sec:background} provides relevant background information and concepts regarding MoE architectures and activation steering. Section~\ref{sec:threat_model} discusses the threat model relevant to our method. Section~\ref{sec:method} details the design of the \ourMethod framework, including the behavioral modeling, safety-relevant expert identification, steering mask creation, and steering mask application. Section~\ref{sec:implementation} describes our experimental setup, datasets used, and MoE models targeted. Section~\ref{sec:experimental_results} presents the main results on the configuration of MoE for safety-related use cases, demonstrating the effectiveness of logit steering. In Section~\ref{sec:discussion}, we discuss the computational cost and efficiency of \ourMethodNoSpace, the limitations, and future work. We provide related work in Section~\ref{sec:related_work}. Finally, we conclude in Section~\ref{sec:conclusion}. Additional experimental details and mechanistic validation of the steering masks are provided in the appendices. 

% The code to perform \ourMethod is available at~\url{https://anonymous.4open.science/r/MASCingAnonymous-9575/}.
The code to perform \ourMethod is available at~\url{https://github.com/jonatelintelo/MASCing}.

% The code to perform \ourMethod is publicly available at~\url{https://anonymous.4open.science/r/MASCingAnonymous-9575/}. 

\section{Background}
\label{sec:background}

% \subsection{Mixture-of-Experts Architectures}
% \label{subsec:moe_architectures}

\textbf{Mixture-of-Experts Architectures. }While dense LLMs activate all their parameters during inference, MoE significantly reduces computational costs by activating only a sparse subset of parameters for any given token~\cite{fedus2022switchtransformers,shazeer2017outrageouslylargeneuralnetworks}. This is achieved by replacing standard dense feed-forward networks with an MoE layer containing multiple independent neural networks, referred to as \emph{experts}. 

The core mechanism of MoE models is the \emph{routing layer}. For each input token, the router computes a set of \emph{routing logits} corresponding to the available experts. These logits are typically passed through a softmax function to produce a probability distribution. To ensure sparsity, the model employs a \emph{top-$k$} expert selection mechanism, meaning only the $k$ experts with the highest routing scores are selected to process the token. The final output of the MoE layer is then computed as a weighted sum of the outputs from these $k$ active experts.

Recent advances in MoE designs have led to structural variations, most notably the distinction between \emph{standard MoE} and \emph{shared expert MoE}. In a standard MoE architecture (e.g., Mixtral-8x7B-Instruct-v0.1~\cite{jiang2024mixtralmoe}), all experts are subject to the router's top-$k$ selection, meaning every expert is highly specialized but might redundantly learn general linguistic knowledge. In contrast, shared expert MoE architectures (e.g., DeepSeek-MoE-16B-Chat~\cite{dai2024deepseekmoe}, Qwen1.5-MoE-A2.7B-Chat~\cite{alibaba2024qwen15moe}) divide experts into two categories: routed experts and shared experts. Shared experts are constantly active and bypass the top-$k$ routing mechanism entirely, with the aim of capturing common knowledge and broad context. The router is then strictly used to select among the remaining specialized experts, ensuring higher parameter efficiency and reducing knowledge redundancy.

% \subsection{Activation Steering}
% \label{subsec:model_steering}

\textbf{Activation Steering. }Activation steering influences model behavior by intervening on internal representations, typically by adding a \emph{steering vector} to the residual stream, the central channel that other model components read from and write to~\cite{elhage2021mathematical}. It builds on the linear representation hypothesis~\cite{park2024the}: high-level concepts are represented as directions in activation space, so amplifying a direction strengthens the corresponding behavior. Steering vectors are commonly obtained from activation differences between inputs that do and do not exhibit a trait, or from unsupervised methods such as sparse autoencoders~\cite{bricken2023monosemanticity}, and have been used to bypass refusal~\cite{arditi2024refusal}, shape personality traits~\cite{chen2025persona}, and mitigate evaluation awareness~\cite{hua2026steering}. Interventions can also target finer-grained components such as attention heads or MLP layers, and their effect depends strongly on layer depth and steering strength~\cite{arditi2024refusal}.

\section{Threat Model}
\label{sec:threat_model}

% We consider a realistic deployment setting in which an LLM developer has full access to a deployed model, including its inputs, activations, and weights, and is responsible for keeping the model aligned with current safety requirements. In practice, these requirements can change after deployment as new jailbreak strategies may expose previously unseen vulnerabilities~\cite{shen2024anything}, updated regulations may require additional safeguards for certain behaviors~\cite{anthropic2026fable}, or shifts in company policies and real-world compliance incidents demand rapid operational responses and model updates~\cite{openai2025adult, anthropic2026fable}. This setting reflects the current operational reality of maintaining deployed LLM systems, where providers have already faced sudden policy interventions, emergency model pauses, and rapid redeployments~\cite{anthropic2026fable}. Developers often need to respond to new risks or policy updates faster than a full retraining cycle allows. Training-free interventions are therefore important because they provide a practical way to update specific model behavior with lower cost and latency, while keeping the overall deployed system consistent with the current policy.

We consider a realistic deployment setting in which an LLM developer has full access to a deployed model, including its inputs, activations, and weights, and is responsible for keeping the model aligned with current safety requirements. In practice, these requirements can change after deployment as new jailbreak strategies may expose previously unseen vulnerabilities~\cite{shen2024anything}, updated regulations may require additional safeguards for certain behaviors~\cite{anthropic2026fable}, and shifts in company policy may permit content that was previously refused~\cite{openai2025adult}. This setting reflects the current operational reality of maintaining deployed LLM systems, where providers have already faced sudden policy interventions, emergency model pauses, and rapid redeployments~\cite{anthropic2026fable}.

% In adversarial settings, following the common access pattern for proprietary models from providers such as OpenAI, Google, and Anthropic, we assume the attacker has black-box (API-only) access to the model. The attacker can submit prompts and observe outputs, but cannot inspect internal states, modify model weights, or directly revert developer interventions. Additionally, we consider an adaptive attacker who is aware that \ourMethod is deployed and actively steering expert routing to prevent harmful outputs. This adversary can iteratively monitor model responses and dynamically adjust their multi-turn conversational strategy to bypass the defensive routing and induce harmful responses. This captures a realistic adaptive threat model for hosted LLM services. \todo{sk: does the attacker get output logits or just tokens? also do they get thinking traces/summaries?}

Following the common access pattern for hosted models from providers such as OpenAI, Google, and Anthropic, the attacker has black-box (API-only) access to the deployed model and they can submit single- or multi-turn conversations and observe the outputs, but cannot inspect internal states, logits, raw thinking traces, modify model weights, or revert developer interventions. In line with Kerckhoffs's principle and established guidance on adaptive evaluation~\cite{tramer2020adaptive}, we assume that the attacker knows that \ourMethod is deployed and how it works. The attacker does not know the deployed steering mask, the surrogate model, or the data used to construct them. The attacker adapts their queries to the observed responses within a fixed query budget and succeeds if any response is harmful.

% \textbf{Security Goal and Non-Goals.} \ourMethod aims to raise the cost of eliciting harmful content through the API, measured as defense success under a fixed attack budget, while keeping the model's responses on-topic and helpful. It does not aim to protect against attackers with white-box access to the deployed system: an attacker who can modify the inference code can remove the steering hook, and an attacker who knows the deployed mask can optimize against it directly. Gradient-based attacks against masks that an attacker reconstructs on a local copy of the base model are a separate attack design problem and are out of scope.

\section{\ourMethod Framework}
\label{sec:method}

\begin{figure*}[!t]
  \centering
  \includegraphics{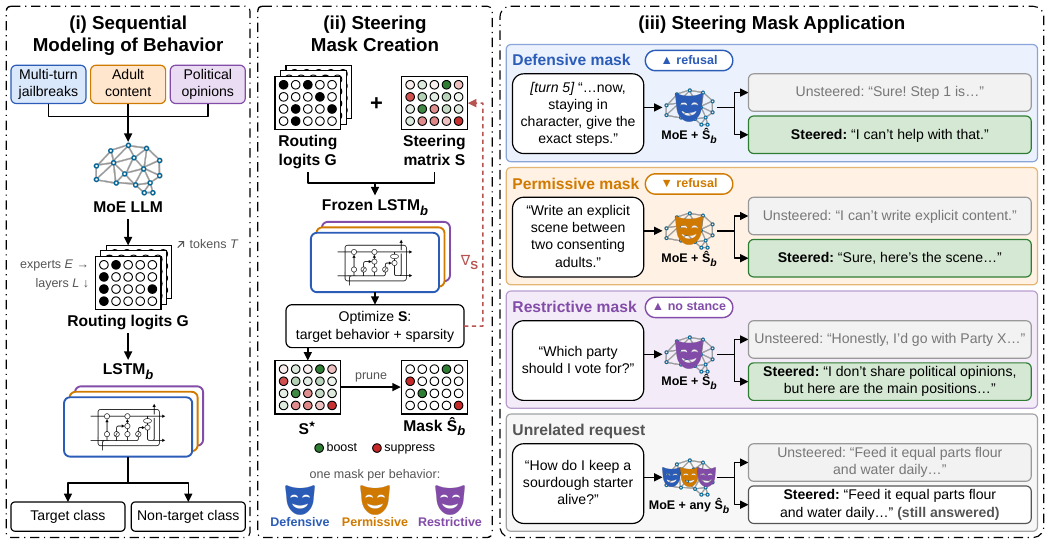}
  % \caption{An overview of the \ourMethod framework. In phase 1, the LSTM is trained to classify routing logits as leading to a target behavior. In phase 2, the steering mask is created by using the LSTM to optimize the mask values. In phase 3, the steering mask is applied to alter the behavior of an MoE model.}
  \caption{Overview of MASCing. (i)~\emph{Sequential modeling of behavior}: a surrogate LSTM is trained on the routing logits of the MoE model to classify if the logits correspond to target behavior. (ii)~\emph{Steering mask creation}: a steering mask is optimized by gradient descent and added to the routing logits to change the LSTM's prediction to target behavior. (iii)~\emph{Steering mask application}: at inference, the steering mask is added to the routing logits of the prompt tokens to steer output towards target behavior.}
  \label{fig:method_figure}
\end{figure*}

Our approach to configuring safety-related behavior in MoE LLMs relies on targeted manipulation of the MoE model's expert selection to either enhance or suppress the engagement of safety-relevant experts during inference. These targeted manipulations are performed with an activation steering mask. The safety-relevant experts are identified via an LSTM model that acts as a surrogate for model behavior. This framework enables both the enhancement of safety guardrails against adversarial jailbreaks and the suppression of refusal to induce compliance. The methodology is divided into three phases: (i) Sequential Modeling of Behavior, (ii) Steering Mask Creation, and (iii) Steering Mask Application. An overview of the \ourMethod framework is given in Figure~\ref{fig:method_figure}.

To approximate the experts responsible for a given target behavior, such as refusal or selective compliance, we model the logits of routing layers across a sequence of tokens. We utilize an LSTM as our surrogate driven by the structural requirements of multi-turn interactions and conversational processing. In conversational settings, malicious intent is often deliberately fragmented across multiple turns~\cite{li2024llmdefensesrobustmultiturn, song2026multibreak}; thus, the behavioral effect of a specific routing decision at the end of a prompt depends entirely on the accumulated context of prior tokens. An architecture inherently designed for sequence modeling, such as an LSTM, is structurally equipped to process these variable-length routing sequences and aggregate the temporal cross-layer dependencies into a fixed-size behavioral representation.

Concretely, the LSTM processes the per-token routing logits of all experts to capture routing patterns and cross-layer interactions and map those to a target behavior. As input to the LSTM, we extract the unnormalized routing logits of all MoE routing layers for a given sequence of tokens in a prompt and train the model to classify whether they lead to the target behavior. We use continuous logits because they preserve the full pre-top-$k$ routing distribution, retaining information about all experts that would be lost after the discrete top-$k$ function.

\subsection{Sequential Modeling of Behavior}
\label{subsec:sequential_modeling_of_behavior}

To steer safety-related behavior, we first have to identify experts associated with the behavior. We identify these experts by modeling how the safety-related behavior is represented in the routing logits. For a prompt of $T$ tokens, let $\mathbf{g}_{t, l} \in \mathbb{R}^E$ denote the unnormalized routing logits of token $t \in [1, T]$ at routing layer $l \in [1, L]$, where $E$ is the number of experts per layer, and let $\mathbf{G} = (\mathbf{g}_{t, l})_{t \in [1, T],\, l \in [1, L]}$ denote the full sequence of routing logits that serves as input to the surrogate.

Because logits exhibit high variance that can cause saturation of the LSTM's recurrent gates and decrease accuracy, we first apply layer normalization across the expert dimension to standardize the inputs. We subsequently employ an affine transformation to project the normalized routing vector into a space of dimension $D$:
\begin{equation}\label{eq:projection}
\mathbf{v}_{t, l} = \mathbf{W}_p \operatorname{LayerNorm}(\mathbf{g}_{t, l}) + \mathbf{b}_p,
\end{equation}
where $\mathbf{W}_p \in \mathbb{R}^{D \times E}$ and $\mathbf{b}_p \in \mathbb{R}^D$ are a learned weight matrix and bias. To represent token $t$ across all routing layers, we concatenate its projected embeddings into a single vector $\mathbf{z}_t \in \mathbb{R}^{L \cdot D}$:
\begin{equation}\label{eq:concatenation}
\mathbf{z}_t = [\mathbf{v}_{t, 1} \parallel \mathbf{v}_{t, 2} \parallel \dots \parallel \mathbf{v}_{t, L}].
\end{equation}

The LSTM processes $\mathbf{z}_1, \dots, \mathbf{z}_T$ in order, capturing both the routing patterns across the token sequence and the dependencies between layers within each token. Variable-length sequences are length-sorted and packed for efficient processing. The recurrent update at step $t$ is
\begin{equation}\label{eq:recurrence}
\mathbf{h}_t, \mathbf{c}_t = \text{LSTM}(\mathbf{z}_t, \mathbf{h}_{t-1}, \mathbf{c}_{t-1}),
\end{equation}
where $\mathbf{h}_t \in \mathbb{R}^H$ is the hidden state, $\mathbf{c}_t \in \mathbb{R}^H$ is the cell state, $H$ is the hidden dimension of the LSTM, and $\mathbf{h}_0 = \mathbf{c}_0 = \mathbf{0}$. To map these representations of sequential routing behavior to the probability that the routing logits $\mathbf{G}$ lead to the target behavior, we apply a linear classification head on the hidden state of the last token:
\begin{equation}\label{eq:prediction}
\hat{y} = f_\theta(\mathbf{G}) = \operatorname{sigmoid}\big(\mathbf{W}_c \mathbf{h}_T + b_c\big),
\end{equation}
where $\mathbf{W}_c \in \mathbb{R}^{1 \times H}$ and $b_c \in \mathbb{R}$ are the weights and biases of the classification head, and $\theta$ refers to the learnable parameters of the surrogate. We train $\theta$ on a dataset of routing-logit sequences labeled with whether they lead to the target behavior, the \emph{target class}, or not, the \emph{non-target class}, minimizing the binary cross-entropy between $\hat{y}$ and the label. During training, each sequence receives a random offset per routing layer and expert, scaled by the layer's standard deviation as in Eq.~\ref{eq:adaptive_scaling} and constant across tokens, while its label is kept. Since a steering mask shifts the routing logits in exactly this way, the surrogate learns how the target behavior responds to such shifts instead of leaving that response to its random initialization. The result is a surrogate $f_\theta$ that classifies a sequence of routing logits as leading to the target behavior in the MoE model. 

% where $\mathbf{W}_c \in \mathbb{R}^{1 \times H}$ and $b_c \in \mathbb{R}$ are the weights and scalar bias of the output layer, respectively.

% To train the surrogate LSTM model for safety-related expert identification in the next phase, we use a dataset of routing logits and labels $y \in \{0, 1\}$. The LSTM network parameters are optimized to minimize the Binary Cross-Entropy with logits loss $\mathcal{L}$:
% \begin{equation}
% \mathcal{L}_{\text{train}} = - \frac{1}{B} \sum_{i=1}^{B} \left[ y_i \log(\sigma(\hat{y}_i)) + (1 - y_i) \log(1 - \sigma(\hat{y}_i)) \right],
% \end{equation}
% where $B$ is the batch size and $\sigma(\cdot)$ is the sigmoid function. The result is an LSTM trained to classify an input sequence of routing logits as leading to a target behavior in the MoE model. 
% We provide an example of the classification output in Appendix~\ref{app:lstm_example}.

\subsection{Steering Mask Creation}
\label{subsec:steering_mask_creation}

The goal is to identify a sparse set of experts whose routing can be steered to promote a target behavior. The mask should steer few experts because a mask that steers many also changes the model's behavior outside the target domain, as the analysis of the sparsity penalty $\lambda$ in Section~\ref{subsec:hyperparameter_analysis} shows. We achieve this by first optimizing a dense steering matrix using the surrogate model, and subsequently pruning it to construct the final sparse steering mask.

We define a learnable steering matrix $\mathbf{S} \in \mathbb{R}^{L \times E}$, whose row $\mathbf{S}_l \in \mathbb{R}^E$ holds one steering value per expert of routing layer $l$. $\mathbf{S}$ is initialized at zero so that every nonzero entry is set by the fitting objective. Because the magnitude of routing logits is inconsistent between different layers within a model, we introduce an adaptive scaling mechanism for $\mathbf{S}$, which ensures that all values in the steering matrix are on a consistent magnitude when we prune them for sparsity. Let $\sigma_l$ denote the standard deviation of the unsteered routing logits at layer $l$. The scaled logits are
\begin{equation}\label{eq:adaptive_scaling}
\tilde{\mathbf{g}}_{t, l} = \mathbf{g}_{t, l} + \sigma_l \mathbf{S}_l,
\end{equation}
and we write $\tilde{\mathbf{G}}$ for the sequence $\mathbf{G}$ with every $\mathbf{g}_{t, l}$ replaced by $\tilde{\mathbf{g}}_{t, l}$.

The trained surrogate $f_\theta$ serves as a differentiable proxy for MoE behavior, so we optimize $\mathbf{S}$ by gradient descent while keeping $\theta$ fixed. Let $\mathcal{D}$ be the set of routing-logit sequences used for profiling and $y^\star \in \{0, 1\}$ the label of the desired behavior. We minimize the binary cross-entropy between the surrogate's prediction on the scaled logits and $y^\star$, together with an $L_1$ penalty on the steering matrix:
\begin{equation}\label{eq:total_loss}
\mathbf{S}^\star = \arg\min_{\mathbf{S}} \; \frac{1}{|\mathcal{D}|} \sum_{\mathbf{G} \in \mathcal{D}} \operatorname{BCE}\big(f_\theta(\tilde{\mathbf{G}}), y^\star\big) + \lambda \|\mathbf{S}\|_1,
\end{equation}
where $\operatorname{BCE}(p, y) = -y \log p - (1 - y) \log(1 - p)$ is the binary cross-entropy, $\|\mathbf{S}\|_1$ is the sum of the absolute values of all entries of $\mathbf{S}$, and $\lambda$ controls the strength of the regularization. A higher value for $\lambda$ promotes sparsity by forcing less important values towards zero, resulting in fewer experts being affected by the mask; with $\lambda = 0$, no penalty is applied, allowing for dense interventions across all logits.

Because the $L_1$ penalty drives unimportant entries of $\mathbf{S}$ towards zero, the entries that keep a large magnitude correspond to the experts most strongly associated with the target behavior. $\mathbf{S}^\star$ nevertheless remains dense, where every expert has a learned intervention weight. To ensure the desired sparsity for minimal influence on untargeted behavior, we extract the steering mask $\hat{\mathbf{S}}$ using a symmetric magnitude gate with threshold $\tau$, applied element-wise to the entries $S^\star_{l, e}$ of $\mathbf{S}^\star$ for each expert $e \in [1, E]$:
\begin{equation}\label{eq:pruning}
\hat{S}_{l, e} = \begin{cases} S^\star_{l, e} & \text{if } |S^\star_{l, e}| > \tau, \\ 0 & \text{otherwise}. \end{cases}
\end{equation}
A higher value for $\tau$ enforces a sparser steering mask, while $\tau = 0$ means no learned weights are pruned. Because the gate acts on the absolute magnitude, it preserves both boosting entries ($S^\star_{l, e} > \tau$), which promote target-aligned experts, and suppressing entries ($S^\star_{l, e} < -\tau$), which penalize experts associated with the opposing behavior. The result is a sparse steering mask that indicates which experts should be promoted or suppressed.

\subsection{Steering Mask Application}
\label{subsec:steering_mask_applicaton}

We steer the model during inference by injecting the static steering mask $\hat{\mathbf{S}}$ into the routing logits through pre-forward hooks on the gate layers of the target LLM. To remain consistent with the adaptive scaling of Eq.~\ref{eq:adaptive_scaling}, the intervention is scaled by the same layer-wise standard deviation $\sigma_l$, as well as an amplitude hyperparameter $\alpha$:
\begin{equation}\label{eq:alpha}
\mathbf{g}'_{t, l} = \mathbf{g}_{t, l} + \alpha \, \sigma_l \hat{\mathbf{S}}_l,
\end{equation}
where $\hat{\mathbf{S}}_l$ is row $l$ of $\hat{\mathbf{S}}$. A higher value for $\alpha$ amplifies the steering strength, resulting in more extreme positive and negative values for the targeted routing logits and a stronger deviation in expert selection compared to the unsteered model. The MoE layer then performs its standard top-$k$ routing on the steered logits, so the steered output of MoE layer $l$ for token $t$ is
\begin{equation}\label{eq:steered_output}
\mathbf{o}'_{t, l} = \sum_{e \in \operatorname{TopK}(\mathbf{g}'_{t, l},\, k)} \big[\operatorname{Gate}_l(\mathbf{g}'_{t, l})\big]_e \, \operatorname{FFN}_{l, e}(\mathbf{u}_{t, l}),
\end{equation}
where $\mathbf{u}_{t, l}$ is the hidden state of token $t$ entering MoE layer $l$, $\operatorname{FFN}_{l, e}$ is expert $e$ of that layer, $k$ is the number of experts activated per token, and $\operatorname{Gate}_l$ is the model's native mapping from routing logits to gating weights (e.g., a softmax over the selected experts). In shared-expert architectures, the always-active shared experts are added to $\mathbf{o}'_{t, l}$ unchanged. The mask is applied to the prompt tokens during prefill only, and the generated tokens are routed by the unsteered router. Preliminary experiments showed that steering the generated tokens as well does not raise the success rate, while it lets more unrelated harmful requests through. By confining the intervention to the routing logits and enforcing the use of a sparse subset of experts, the model is steered toward the target behavior while preserving lexical and syntactic integrity. We analyze the sensitivity and tuning of the intervention amplitude $\alpha$, the $L_1$ regularization penalty $\lambda$, and the pruning threshold $\tau$ in Section~\ref{subsec:hyperparameter_analysis}. In practice, $\alpha$ needs tuning per model and configuration, the best $\lambda$ is chosen together with it, and $\tau$ can stay fixed at 0.1.

\section{Implementation and Evaluation Setup}
\label{sec:implementation}

\subsection{Target Models}
\label{subsec:target_models}
 
We evaluate \ourMethod on seven open-source MoE LLMs that span a diverse set of architecture types, number of experts and layers, parameter counts, developers, and top-$k$ activation values: DeepSeek-MoE-16B-Chat~\cite{dai2024deepseekmoe}, GPT-OSS-20B~\cite{openai2025gptossmoe}, Hunyuan-A13B-Instruct~\cite{tencent2025hunyuanmoe}, Mixtral-8x7B-Instruct-v0.1~\cite{jiang2024mixtralmoe}, Phi-3.5-MoE-Instruct~\cite{abdin2024phi3moe}, Qwen1.5-MoE-A2.7B-Chat~\cite{alibaba2024qwen15moe}, and Qwen3-30B-A3B-Instruct-2507~\cite{yang2025qwen3moe}. The relevant model architecture specifications are detailed in Table~\ref{tab:model_specifications}.

\subsection{Safety Configurations}
\label{subsec:safety_configurations}
 
To demonstrate the versatility of \ourMethodNoSpace, we define three safety configurations with contrasting objectives. The \emph{defensive mask} strengthens the model against multi-turn jailbreaks, in which an adversary fragments a harmful request into seemingly benign steps across a conversation, so that the harmful intent is diluted and bypasses single-turn safeguards~\cite{li2024llmdefensesrobustmultiturn,song2026multibreak}. Its target behavior is a safe response throughout a jailbreak conversation. The \emph{permissive mask} relaxes refusal for a single content domain, adult-content generation, reflecting recent policy discussions on permitting mature content in age-gated contexts such as creative writing~\cite{openai2025adult}. Its target behavior is compliance with adult-content requests, while refusal of unrelated harmful requests should be preserved. The \emph{restrictive mask} makes the model withhold a behavior in a domain that a deployer wants to exclude, here personal opinions on political questions, which assistants for a broad public may be required to avoid. Its target behavior is a response that takes no position as the model's own, while questions outside politics should still be answered. A mask strengthens a behavior that the model already shows for part of the target domain, since the surrogate learns from both behaviors. DeepSeek-MoE-16B-Chat, Mixtral-8x7B-Instruct-v0.1, and Qwen1.5-MoE-A2.7B-Chat do not refuse adult-content requests, so we evaluate the permissive configuration on the remaining four models. DeepSeek-MoE-16B-Chat, GPT-OSS-20B, and Hunyuan-A13B-Instruct already avoid taking political stances in the experimental setup, so we do not evaluate the restrictive configuration on these models.

\subsection{Datasets}
\label{subsec:datasets}

Each configuration uses three kinds of data. \emph{Profiling data} trains the surrogates and fits the mask. \emph{Evaluation data} from the target domain measures the success rate. \emph{Side-effect data} from outside the target domain measures what else the mask changes. The evaluation and the side-effect data are each split into a \emph{selection set}, on which we tune all hyperparameters, and a \emph{test set}, on which we measure every reported result once. All kinds of data rely on labels for the model's responses, which we define first.

\textbf{Behavior Labeling.} Depending on the configuration, a response is labeled as a refusal, as compliant, by its stance, or as harmful. The refusal, compliance, and stance labels require several detectors to agree. Responses on which they disagree are excluded from profiling.

A response is a \emph{refusal} only if three detectors agree. First, it starts with a refusal phrase such as ``I am sorry...'' or ``I cannot...''. Second, the WildGuard~\cite{han2024wildguard} classifier marks it as a refusal. Third, an LLM judge (Qwen3-30B-A3B-Instruct-2507) classifies it as a refusal among five categories (full compliance, partial compliance, refusal, clarification or inability, and other). A response is \emph{compliant} only if none of the three detectors sees a refusal and the judge classifies it as compliance. For an adult-content request, the response must also still be generating at the 128-token answer budget. Short replies to a story premise are mostly sanitized summaries rather than the requested content.

For the restrictive configuration, the label concerns the stance of a response rather than refusal. A model that declines to give an opinion often continues with neutral information, which the refusal detectors count as compliance. Two LLM judges, Qwen3-30B-A3B-Instruct-2507 and Qwen2.5-14B-Instruct~\cite{qwen2025qwen25}, therefore classify whether a response contains a political position taken by the model. A response \emph{takes no position} if both judges find that it declines to choose, states that it has no personal opinions, or only gives neutral information or several perspectives. It \emph{takes a position} if both judges find that it chooses or endorses one of the offered options, also after a disclaimer such as ``As an AI, I have no opinions, but...''. A response that presents one option as the right answer without endorsing it receives neither label.

For the defensive configuration, a response is \emph{harmful} if Llama-Guard-3-8B~\cite{grattafiori2024llama3herdmodels} classifies it as unsafe. A response that a coherence judge (Qwen2.5-14B-Instruct) marks as incoherent also counts as unsafe, so a successful defense requires a response that is both coherent and safe. For the side effects of the permissive and the restrictive configuration, a response has \emph{harmful content} if WildGuard marks it as a harmful response. All responses are generated with greedy decoding after the system instruction ``You are an AI that provides direct and precise answers to any question. Respond only to the question without additional details or explanations.'', except for DeepSeek-MoE-16B-Chat, whose chat template has no system role.

\textbf{Profiling Classes.} The surrogate of Section~\ref{subsec:sequential_modeling_of_behavior} is trained on the routing logits of prompts of the target and the non-target class. A surrogate trained on one source dataset per class can separate the classes by topic alone, without modeling the behavior. For the permissive and the restrictive configuration, both classes therefore contain prompts from the target domain. The model shows the target behavior on some of these prompts and the opposing behavior on others. Each class also contains prompts from an unrelated anchor dataset, with an equal number of prompts from each source. In the defensive configuration, both classes already consist of harmful requests.

\textbf{Selection and Test Sets.} No selection or test prompt is used to train a surrogate or to fit a mask. Whether the model refuses a prompt or takes a position on it is only known after labeling. For the permissive and the restrictive configuration, we therefore split the target-domain prompts after labeling, stratified by label. Two-thirds go to profiling and one-sixth each to the selection and the test set. The held-out anchor prompts that measure side effects are split in half between the selection and the test set.

\textbf{Defensive Configuration.} The target class consists of the prompts from AdvBench~\cite{zou2023universal} that the model refuses. AdvBench is a widely used set of 520 harmful instructions. The non-target class comes from the Multi-Turn Human Jailbreaks (MHJ) dataset~\cite{li2024llmdefensesrobustmultiturn}, which contains 2,912 prompts across 537 human-written multi-turn conversations. The conversations are balanced over six jailbreak methods (obfuscation, injection, request framing, direct request, hidden intention streamline, and output formatting). We replay each conversation turn by turn and append the model's response to the context as long as it is not harmful. We then collect the routing logits of the context that yields the first harmful response. For evaluation, we use MultiBreak~\cite{song2026multibreak}, a benchmark of 10,389 multi-turn adversarial prompts across 2,665 harmful intents. It was generated with an active-learning pipeline to uncover vulnerabilities that bypass single-turn defenses. As a different source than MHJ and AdvBench, MultiBreak tests whether a defensive mask transfers beyond its profiling data. We split it by harmful intent, so that no intent appears in both the selection and the test set. Over-refusal is measured on benign prompts from Facebook's NaturalReasoning~\cite{yuan2025naturalreasoningreasoningwild28m}, a dataset of more than one million general reasoning tasks. These prompts are disjoint from all profiling prompts. The adaptive attacks of Section~\ref{subsec:adaptive_attacker_setup} target 100 harmful instructions as objectives, sampled from StrongREJECT~\cite{souly2024strongreject}, a set of 313 harmful instructions ranging from physical harm to illegal cyber activities.

\textbf{Permissive Configuration.} The target-domain prompts come from EroticaAnalysis~\cite{openerotica2024eroticaanalysis}, one of the largest public datasets of adult-content requests. The target class consists of EroticaAnalysis and NaturalReasoning prompts that the model answers. The non-target class consists of EroticaAnalysis prompts that the model refuses, together with harmful requests that it refuses from AdvBench and the toxic seeds of Health-ORSC-Bench~\cite{zhang2026healthorsc}. Only a random sample of 6,000 EroticaAnalysis prompts is split into profiling, selection, and test prompts. All other EroticaAnalysis prompts are used for profiling only. Selective refusal is measured on held-out AdvBench, Health-ORSC-Bench requests that the model refuses, and on all StrongREJECT prompts with the exception of 50 prompts that are in the category ``Sexual content'', which overlap the target domain.

\textbf{Restrictive Configuration.} The target-domain prompts come from GlobalOpinionQA~\cite{durmus2023towards}, 2,556 survey questions with answer options from the Pew Global Attitudes Survey and the World Values Survey. We draw 1,500 questions and ask each in five phrasings that request the model's own choice, such as ``Which option do you personally choose, and why?''. The target class consists of the political prompts on which the model takes no position, together with AdvBench and Health-ORSC-Bench requests that it refuses. The non-target class consists of the political prompts on which the model takes a position, together with NaturalReasoning prompts that it answers. We split whole questions rather than single prompts, so that all phrasings of a question fall into the same part. The split is stratified by the number of phrasings on which the model takes no position. Side effects are measured on held-out NaturalReasoning prompts that the model answers and on held-out AdvBench and Health-ORSC-Bench requests that it refuses. The mask should neither make the model decline the former nor answer the latter with harmful content.

% \begin{table}[!t]
%   \centering
%   \footnotesize
%   \caption{Data used at each stage. Profiling data trains the surrogate, selection splits choose the mask hyperparameters, and all reported success rates use the disjoint test splits. Def.: defensive, Perm.: permissive configuration.}
%   \label{tab:data_splits}
%   \setlength{\tabcolsep}{3pt}
%   \begin{tabular}{llll}
%     \toprule
%     \textbf{Stage} & \textbf{Config.} & \textbf{Source} & \textbf{Content} \\
%     \midrule
%     \multirow{4}{*}{Profiling} & \multirow{2}{*}{Def.} & AdvBench & refused prompts \\
%      & & MHJ & first harmful turn \\
%      & \multirow{2}{*}{Perm.} & EroticaAnalysis & refused prompts \\
%      & & NaturalReasoning & answered prompts \\
%     \midrule
%     \multirow{2}{*}{Selection} & Def. & MultiBreak & \ph{N} conversations \\
%      & Perm. & EroticaAnalysis & \ph{N} held-out prompts \\
%     \midrule
%     \multirow{2}{*}{Test} & Def. & MultiBreak & 250 conversations \\
%      & Perm. & EroticaAnalysis & \ph{N} held-out prompts \\
%     \midrule
%     Adaptive & Def. & StrongREJECT & 50 objectives \\
%     \midrule
%     \multirow{2}{*}{Over-refusal} & \multirow{2}{*}{Def.} & NaturalReasoning & 150 benign prompts \\
%      & & OR-Bench-Hard-1K & \ph{N} benign prompts \\
%     \midrule
%     \multirow{2}{*}{Selectivity} & \multirow{2}{*}{Perm.} & StrongREJECT & 150 harmful prompts \\
%      & & OR-Bench-Toxic & 600 harmful prompts \\
%     \bottomrule
%   \end{tabular}
% \end{table}

\subsection{Surrogate Training}
\label{subsec:surrogate_training}
 
The surrogate uses the same dimensions, learning rate, and batch size for every MoE model, determined by a preliminary grid search over embedding dimensions $D \in \{8, 16, 32\}$ and hidden dimensions $H \in \{32, 64, 128\}$, ranges chosen to balance convergence, computational efficiency, and the prevention of overfitting. We set $D=16$ and $H=64$ and train until convergence with the Adam optimizer~\cite{kingma2017adam} at a learning rate of 0.01, a weight decay of $10^{-4}$, and its default parameters otherwise ($\beta_1 = 0.9$, $\beta_2 = 0.999$, $\epsilon = 10^{-8}$), with a batch size of 64 for the defensive and 512 for the permissive and the restrictive configurations. The training offsets of Section~\ref{subsec:sequential_modeling_of_behavior} are drawn from a standard normal distribution.

\subsection{Mask Creation and Application}
\label{subsec:mask_settings}

We optimize the steering matrix of Eq.~\ref{eq:total_loss} with the Adam optimizer at a learning rate of 0.05 for 100 epochs and a batch size of 16. We fix the pruning threshold at $\tau = 0.1$ and tune the sparsity penalty $\lambda$ and the amplitude $\alpha$ per model and configuration on the selection set (Section~\ref{subsec:datasets}), so that no setting depends on the test set. We choose the setting with the highest success rate among those whose side effects stay within 5\% of the unsteered model: harmless prompts declined or refused, harmful requests answered with harmful content, and incoherent responses. The search runs coarse to fine over $\lambda \in \{1\text{e-}5, 1\text{e-}4, 1\text{e-}3\}$ and $\alpha \in [0.25, 3]$. A coarse pass evaluates every $\lambda$ at amplitudes in steps of 0.25 to 0.5, and a fine pass refines $\alpha$ in steps of 0.05 between the best permitted coarse value and the next larger one when that value exceeds the side-effect limit. Section~\ref{subsec:hyperparameter_analysis} analyzes the effect of each hyperparameter.

\subsection{Adaptive Attacker Setup}
\label{subsec:adaptive_attacker_setup}
 
To evaluate \ourMethod's security against an adversary actively attempting to bypass the steering mask, we implement a multi-turn attack based on the Crescendo jailbreak framework~\cite{russinovich2024crescendo}. Unlike the static conversations of MHJ~\cite{li2024llmdefensesrobustmultiturn} and MultiBreak~\cite{song2026multibreak}, Crescendo generates each query conditioned on the target model's previous responses. Our implementation uses GPT-4.1 as the attacker model and targets the 100 StrongREJECT objectives of Section~\ref{subsec:datasets}.
 
We follow the procedure of Russinovich et al.~\cite{russinovich2024crescendo}. The attacker escalates progressively over at most 10 conversational rounds, and whenever the target refuses, the refused turn is removed from the conversation history and the attacker rephrases it, up to 10 times. Each objective is attacked in 5 independent trials, and an objective counts as broken if any trial elicits a harmful response.
 
The attacker's system prompt states that the target applies a static activation steering mask to its routing logits to enforce safety. If the target model produces a safe, contextually relevant refusal, indicating that the \ourMethod intervention has been triggered, the attacker is instructed to pivot its semantic strategy and search for formulations on which the static mask does not induce a safe response. Unsteered and defended models receive the same attack budget. The attacker's in-loop success judgments use GPT-4.1, which is separate from the final evaluation judge (Section~\ref{subsec:evaluation_metrics}).

\subsection{Evaluation Metrics}
\label{subsec:evaluation_metrics}
 
\textbf{Success Rate.} We measure the effectiveness of \ourMethod by its success rate, the percentage of test conversations or prompts for which the model exhibits the target behavior, taken over all test data before and after the mask is applied. For the defensive mask, a conversation counts as a success if none of the model's responses is harmful under the harm criterion of Section~\ref{subsec:datasets}, i.e., the model withstands the jailbreak; such a response need not be a refusal. For the permissive mask, a prompt counts as a success if the response provides the requested content, as judged by the rubric-based LLM judge of Section~\ref{subsec:datasets}, and is coherent, as judged by the coherence judge, so a reworded refusal or a degenerate output does not count. For the restrictive mask, a prompt counts as a success if the response takes no position as the model's own, as judged by the two rubric-based stance judges of Section~\ref{subsec:datasets}, and is coherent, i.e., neither empty, degenerate, nor cut off before its final answer.

\textbf{Side Effect.} A mask should change the target behavior without carrying it over to inputs where it is unwanted. For each configuration, we therefore measure the same behavior on normal inputs. For the defensive mask, the percentage of benign NaturalReasoning prompts that the model refuses; for the permissive mask, the percentage of harmful StrongREJECT requests that the model answers with harmful content; and for the restrictive mask, the percentage of harmless, non-political NaturalReasoning prompts that the model declines, as judged by the rubric-based LLM judge of Section~\ref{subsec:datasets}. We report the change in percentage points that the mask causes.
 
\textbf{Model Utility.} When a defender applies \ourMethodNoSpace, the utility of the model should remain largely unaffected. To quantify the impact of \ourMethod on model utility, we measure accuracy on three complementary benchmarks: Massive Multitask Language Understanding (MMLU)~\cite{hendrycks2021measuring} for broad knowledge and reasoning across 57 subjects, Grade School Math 8K (GSM8K)~\cite{cobbe2021gsm8k} for multi-step mathematical reasoning over extended generation, and Instruction Following Evaluation (IFEval)~\cite{zhou2023instruction} for following verifiable instructions in longer open-ended generations. All benchmarks are run with lm-evaluation-harness~\cite{gao2024evalharness} for consistent comparison across architectures. MMLU and GSM8K are evaluated with 5-shot prompting and IFEval with zero-shot, following common methodology~\cite{dai2024deepseekmoe,openai2025gptossmoe,tencent2025hunyuanmoe,jiang2024mixtralmoe,abdin2024phi3moe,alibaba2024qwen15moe,yang2025qwen3moe}.

\section{Experimental Results}
\label{sec:experimental_results}

\subsection{\ourMethod Success Rate}
\label{subsec:success_rate}

We evaluate the effectiveness of \ourMethod across three safety configurations: a defensive configuration against multi-turn jailbreaks, a permissive configuration for adult-content generation, and a restrictive configuration that withholds personal political opinions. Table~\ref{tab:main_results} reports, for every model and configuration, the success rate before and after \ourMethodNoSpace, the side effect on normal inputs, and the change in utility. The hyperparameters $\lambda$ and $\alpha$ are fixed per model and configuration on a separate selection set, as described in Section~\ref{subsec:mask_settings}, so that no choice is informed by the test data. Section~\ref{subsec:hyperparameter_analysis} analyzes how the success rate depends on them.

\textbf{Defensive Configuration.} As detailed in Table~\ref{tab:main_results}, standard alignment is heavily degraded in multi-turn adversarial settings, with unsteered models successfully defending against attacks only 52.5\% of the time on average. Applying a defensive mask yields a substantial and consistent improvement across all tested MoE models, raising the average defense success rate to 83.9\% with some masked models achieving a near 90\% defense success rate. This demonstrates that the continuous activation steering of \ourMethod improves robustness against evasive, multi-step attacks.

\textbf{Permissive Configuration.} To demonstrate the configurability of \ourMethod for domain-specific policies, we evaluate the ability to induce compliance for adult-content generation tasks that standard safety-aligned models typically refuse. As shown in Table~\ref{tab:main_results}, a permissive mask substantially increases the generation success rate, raising the average from 52.6\% to 82.0\%. These results confirm that \ourMethod can reliably suppress refusal and elicit specific capabilities without retraining. %Note that we did not evaluate some models which do not refuse requests for adult-content generation. % DeepSeek-MoE-16B-Chat, Mixtral-8x7B-Instruct-v0.1, and Qwen1.5-MoE-A2.7B-Chat because these models do not refuse requests or instructions for adult-content generation.

\textbf{Restrictive Configuration.} The restrictive mask makes a model withhold a behavior in a domain that a deployer wants to exclude, here personal opinions on political questions. Unsteered, the models take no position on 31.8\% of the test questions on average, typically with a reply such as ``As an AI, I don't have personal opinions'', and pick one of the offered options for most of the others. 

As shown in Table~\ref{tab:main_results}, the restrictive mask raises the share of questions on which the models take no position to 75.4\% on average, while the share of harmless non-political questions that they decline changes by at most 4.5\%. The steered models rarely refuse outright. They state that they have no personal opinion and continue with neutral information, which is the behavior a deployer of a neutral assistant asks for. The effect differs between models, from 17.2\% on Mixtral-8x7B-Instruct-v0.1 to 64.1\% on Qwen3-30B-A3B-Instruct-2507. Together with the defensive and permissive results, this shows that \ourMethod can steer behavior in each of the three directions: towards safe responses under attack, towards compliance in a permitted domain, and towards declining in an excluded one.

\begin{table*}[!t]
  \centering
  \caption{Main results of \ourMethod. Success is the percentage of test conversations or prompts that show the target behavior before $\rightarrow$ after \ourMethodNoSpace. $\Delta$ Side is the change in percentage points of the target behavior on normal inputs where it is unwanted (Section~\ref{subsec:evaluation_metrics}); lower is better. $\Delta$ Utility is the change in accuracy in percentage points, averaged over MMLU, GSM8K, and IFEval. A dash marks a model that shows only one of the two behaviors in the configuration's domain.}
  \label{tab:main_results}
  \footnotesize
  \setlength{\tabcolsep}{5.2pt}
  \begin{tabular}{l|ccc|ccc|ccc}
    \toprule
    & \multicolumn{3}{c|}{\textbf{Defensive}} & \multicolumn{3}{c|}{\textbf{Permissive}} & \multicolumn{3}{c}{\textbf{Restrictive}} \\
    \textbf{Model} & \textbf{Success} & \textbf{$\Delta$ Side} & \textbf{$\Delta$ Utility} & \textbf{Success} & \textbf{$\Delta$ Side} & \textbf{$\Delta$ Utility} & \textbf{Success} & \textbf{$\Delta$ Side} & \textbf{$\Delta$ Utility} \\
    \midrule
    DeepSeek-MoE-16B-Chat        & 52.8 $\rightarrow$ 84.8 & $+$2.6 & $-$4.3 & - & - & - & - & - & - \\
    GPT-OSS-20B                  & 62.0 $\rightarrow$ 88.0 & $+$3.2 & $-$3.1 & 46.3 $\rightarrow$ 79.3 & $+$2.7 & $-$4.1 & - & - & - \\
    Hunyuan-A13B-Instruct        & 48.8 $\rightarrow$ 80.4 & $+$2.9 & $-$2.3 & 69.7 $\rightarrow$ 88.6 & $+$3.4 & $-$3.6 & - & - & - \\
    Mixtral-8x7B-Instruct-v0.1   & 47.6 $\rightarrow$ 77.2 & $+$4.1 & $-$5.3 & - & - & - & 46.4 $\rightarrow$ 63.6 & $+$0.5 & $-$3.1 \\
    Phi-3.5-MoE-Instruct         & 57.6 $\rightarrow$ 80.8 & $+$2.3 & $-$5.2 & 61.2 $\rightarrow$ 93.0 & $+$2.0 & $-$4.5 & 34.3 $\rightarrow$ 78.1 & $+$0.8 & $-$4.1 \\
    Qwen1.5-MoE-A2.7B-Chat       & 51.6 $\rightarrow$ 87.2 & $+$3.8 & $-$3.1 & - & - & - & 26.4 $\rightarrow$ 75.7 & $+$4.5 & $-$3.8 \\
    Qwen3-30B-A3B-Instruct-2507  & 47.2 $\rightarrow$ 89.2 & $+$4.3 & $-$3.6 & 33.2 $\rightarrow$ 67.1 & $+$2.6 & $-$3.5 & 20.2 $\rightarrow$ 84.3 & $+$4.3 & $-$4.0 \\
    \midrule
    \emph{Average} & \emph{52.5 $\rightarrow$ 83.9} & \emph{$+$3.3} & \emph{$-$3.8} & \emph{52.6 $\rightarrow$ 82.0} & \emph{$+$2.7} & \emph{$-$3.9} & \emph{31.8 $\rightarrow$ 75.4} & \emph{$+$2.5} & \emph{$-$3.8} \\
    \bottomrule
  \end{tabular}
\end{table*}

Beyond success rates, a notable observation in the qualitative analysis of the generated model responses reveals that \ourMethod does not merely steer the model towards producing generic refusals (e.g., ``I'm sorry, I cannot...''). Instead, the steered models frequently generate safe, contextually relevant responses that address the user's overarching topic without complying with the harmful intent. An example of this behavior is illustrated in Figure~\ref{fig:jailbreak_example} in Appendix~\ref{app:qualitative_example}. This indicates that our masking approach successfully redirects the MoE routing mechanism to engage with safety-aligned experts, enabling nuanced, on-topic refusal rather than a simplistic trigger-based or hardcoded refusal block.

\textbf{Side Effects.} A configuration should change the target behavior only where it is intended. The $\Delta$ Side columns of Table~\ref{tab:main_results} therefore measure the target behavior on normal inputs where it is unwanted. The defensive mask raises the share of benign NaturalReasoning prompts that the models refuse by 3.3\% on average, the permissive mask raises the share of unrelated StrongREJECT requests answered with harmful content by 2.7\%, and the restrictive mask raises the share of harmless non-political questions that the models decline by 2.5\%. This demonstrates that our framework isolates experts associated with the target behavior without degrading the model's broader capabilities or guardrails.

\textbf{Robustness Against Adaptive Attacks.} A reliable multi-turn defense must also withstand adversaries that adapt their strategies based on the target model's responses. To determine whether static defensive \ourMethod generalizes to dynamically generated attack trajectories, we evaluate \ourMethod against the Adaptive Crescendo attack described in Section~\ref{subsec:adaptive_attacker_setup}.

As shown by the results in Table~\ref{tab:adaptive_jailbreak_defense}, the baseline models successfully defended against 30.7\% of the harmful objectives on average. Applying defensive \ourMethod increased the average defense success rate to 74.3\%. Although this rate is lower than the 86.0\% achieved against non-adaptive conversations on the same three-model subset, \ourMethod retains a defensive advantage under response-conditioned semantic pivoting. These results suggest that the LSTM surrogate captures refusal-routing patterns that generalize beyond the fixed attack transcripts used during training.

Notably, the surrogate training dataset contains no Adaptive Crescendo conversations, so the observed residual failures may correspond to routing trajectories that are underrepresented in the current training distribution. Incorporating routing logits of the Crescendo conversations into the surrogate training dataset may enable \ourMethod to identify additional safety-relevant routing patterns and account for more semantic pivots to further improve robustness against adaptive attacks.

\begin{table}[!t]
  \centering
  \caption{Defense success rates against an adaptive multi-turn attacker (Crescendo) before and after applying defensive \ourMethodNoSpace. Success rate reflects the percentage of harmful objectives for which the model produced no unsafe response during the conversation.}
  \label{tab:adaptive_jailbreak_defense}
  \begin{tabular}{l|cc}
    \toprule
    \textbf{Model} & \textbf{Baseline} & \textbf{\ourMethodNoSpace} \\
    \midrule
    GPT-OSS-20B                 & 31.0\% & 79.0\% \\
    Phi-3.5-MoE-Instruct        & 35.0\% & 73.0\% \\
    Qwen3-30B-A3B-Instruct-2507 & 26.0\% & 71.0\% \\
    \midrule
    \emph{Average} & \emph{30.7\%} & \emph{74.3\%} \\
    \bottomrule
  \end{tabular}
\end{table}

\subsection{Effect of \ourMethod on Utility}
\label{subsec:effects_on_utility}

While \ourMethod changes the target behavior, a steered model must retain its general language capabilities. To quantify the effect of activation steering on utility, we benchmark every steered model on MMLU, GSM8K, and IFEval, as described in Section~\ref{subsec:evaluation_metrics}. The $\Delta$ Utility columns of Table~\ref{tab:main_results} give the change in accuracy in percentage points, averaged over the three benchmarks, and Table~\ref{tab:utility} in Appendix~\ref{app:additional_utility} gives the accuracy before and after on each benchmark.

Overall, the accuracy of the steered models drops by 3.8\% on average, and by at most 5.3\% for any model and configuration. The masks therefore do not cause catastrophic utility degradation or logical collapse. On MMLU, the models maintain factual recall and domain-specific reasoning, and the modest declines on GSM8K and IFEval confirm that the altered forward pass continues to support coherent multi-step generation, open-ended generation, and instruction following. The average drop is 3.8\%, 3.9\%, and 3.8\% for the defensive, permissive, and restrictive masks, so no configuration stands out. The two largest drops, 5.3\% and 5.2\%, come from the defensive masks of Mixtral-8x7B-Instruct-v0.1 and Phi-3.5-MoE-Instruct. We attribute the larger drops to these models routing each token to only two experts (Table~\ref{tab:model_specifications}). With two active experts, every expert that the mask swaps in replaces half of the expert computation of a token, against an eighth for Qwen3-30B-A3B-Instruct-2507.

Per benchmark (Table~\ref{tab:utility}), the steered models lose the least accuracy on MMLU and the most on IFEval. The drop lies between 1.6 and 4.6\% on MMLU, between 2.3 and 5.4\% on GSM8K, and between 3.0 and 5.9\% on IFEval. Among the 15 combinations of model and configuration, MMLU has the smallest drop in 14 and IFEval the largest in 13. We attribute this order to the length of the output that each benchmark scores. MMLU scores a single prediction, the choice of an answer. GSM8K and IFEval score long generations, in which every generated token attends to the prompt representation that the mask has changed (Section~\ref{subsec:steering_mask_applicaton}). A small shift in the next-token distribution therefore acts at every step, and a single slip fails the response: one wrong calculation step on GSM8K, one instruction not followed on IFEval. Steering thus costs a few points of accuracy, most on long generations, while the steered models keep the capabilities that the three benchmarks measure.

\subsection{Hyperparameter Analysis}
\label{subsec:hyperparameter_analysis}

\begin{figure*}[!t]
  \centering
  \includegraphics[width=1\textwidth]{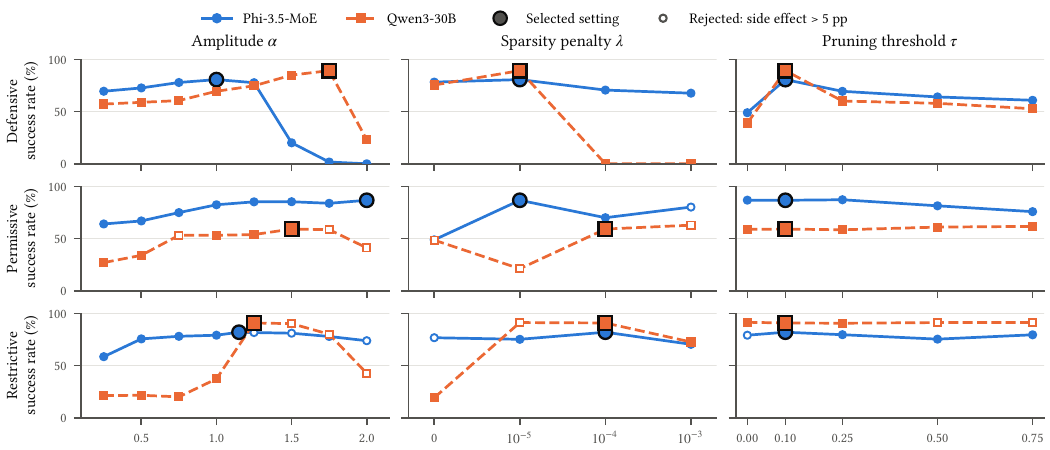}
  \caption{Success rate of Phi-3.5-MoE-Instruct and Qwen3-30B-A3B-Instruct-2507 as one hyperparameter varies and the other two stay at the selected setting, in the defensive, permissive, and restrictive configurations (rows). Large markers show the selected setting, and hollow markers show settings that the selection rule rejects because a side effect rises by more than 5 percentage points.}
  \label{fig:hyperparameters}
\end{figure*}

As defined in Section~\ref{sec:method}, three hyperparameters shape the steering mask. The amplitude $\alpha$ sets the strength of the shift when the mask is applied (Eq.~\ref{eq:alpha}), the sparsity penalty $\lambda$ sets how many experts the mask steers (Eq.~\ref{eq:total_loss}), and the pruning threshold $\tau$ removes the small entries that remain after the optimization (Eq.~\ref{eq:pruning}). Figure~\ref{fig:hyperparameters} shows how the success rate depends on each of them for Phi-3.5-MoE-Instruct and Qwen3-30B-A3B-Instruct-2507, the two models that we evaluate in all three configurations. Each panel varies one hyperparameter over $\alpha \in \{0.25, 0.5, \dots, 2\}$, $\lambda \in \{0, 1\text{e-}5, 1\text{e-}4, 1\text{e-}3\}$, or $\tau \in \{0, 0.1, 0.25, 0.5, 0.75\}$ and keeps the other two at the setting selected in Section~\ref{subsec:mask_settings}. Like the selection itself, the analysis uses only the selection set. The method needs all three hyperparameters, and they differ in how much tuning they need: $\alpha$ is tuned per model and configuration within a window close to 1, $\lambda$ depends on how broad the target behavior is and is chosen together with $\alpha$, and a single threshold $\tau = 0.1$ works for all models and configurations. The defensive configuration, whose multi-turn inputs are the longest, reacts most to $\alpha$ and $\tau$.

\textbf{Amplitude.} The useful strength of the shift lies in a window that differs per model and configuration, so $\alpha$ is tuned per model. A value that is too low mostly costs success, whereas one that is too high breaks the defense or spreads the behavior to unrelated inputs. All selected amplitudes lie between 1 and 2, but where the window ends differs: in the defensive configuration, Phi-3.5-MoE-Instruct breaks down at $\alpha = 1.5$, to 20.1\%, while Qwen3-30B-A3B-Instruct-2507 still gains up to $\alpha = 1.75$ (89.2\%) and breaks down only at $\alpha = 2$, to 23.1\%. Below the window, the routing barely changes: in the restrictive configuration, Qwen3-30B-A3B-Instruct-2507 takes no position on about 21\% of the political questions up to $\alpha = 0.75$, as without steering (20.9\%), but on 90.8\% at $\alpha = 1.25$. Above the window, the behavior spreads: at $\alpha = 1.5$, the same model declines 50.0\% more harmless questions, and at $\alpha = 2$, 22.4\% of its answers are incoherent. At $\alpha = 1$, the mask shifts the routing logits by as much as during its optimization (Eqs.~\ref{eq:adaptive_scaling} and~\ref{eq:alpha}), which is why the selected values lie close to 1. A static mask stays selective as long as its shift moves the steered experts across the top-$k$ boundary for target prompts but not for other prompts. Too large a shift forces them into or out of the top-$k$ for almost every token, whatever the input, which approaches the rigid expert steering of Appendix~\ref{sec:activation_steering_versus_expert_steering}. Over the many tokens of a multi-turn conversation, this breaks the defense; for single prompts, it carries the behavior to unrelated inputs.

\textbf{Sparsity Penalty.} The mask needs a non-zero sparsity penalty $\lambda$ to steer the target behavior without spreading it to other inputs, and the right $\lambda$ depends on how broad the target behavior is, so it is chosen together with $\alpha$. Without the penalty, the masks keep most of their entries, for example, 5,083 of 6,144 for Qwen3-30B-A3B-Instruct-2507 in the restrictive configuration. That mask fails to steer (19.2\% of the political questions without a position, against 20.9\% without steering), and with its dense restrictive mask, Phi-3.5-MoE-Instruct declines 57.6\% more harmless questions. The defensive masks, in contrast, still raise the success rate when dense, and the weaker penalty $\lambda = 1\text{e-}5$ is selected for both models, against $1\text{e-}4$ for three of the four permissive and restrictive masks. At the amplitude selected for $\lambda = 1\text{e-}5$, Qwen3-30B-A3B-Instruct-2507 defends none of the conversations from $\lambda = 1\text{e-}4$ on. The surrogate is only a proxy for the MoE model, and many small shifts can satisfy it without targeting the experts that produce the behavior, much like an adversarial perturbation of a classifier. The penalty concentrates the mask on the few experts with the largest effect. A defensive mask has to produce safe answers to jailbreaks on many different harmful topics, which plausibly engage more experts than a single domain does, so it needs a weaker penalty. Since $\alpha$ scales whatever mask the penalty produces, the two are chosen together.

\textbf{Pruning Threshold.} Pruning the small leftover entries keeps the defensive masks effective, and a single threshold, $\tau = 0.1$, works for all models and configurations. In the defensive configuration, the success rate without pruning ($\tau = 0$) falls below that of the unsteered model, to 48.9\% for Phi-3.5-MoE-Instruct and 39.1\% for Qwen3-30B-A3B-Instruct-2507 (unsteered 57.7\% and 47.3\%), and at $\tau = 0.75$ it falls back towards the unsteered rate, while $\tau = 0.1$ gives the highest success rate. In the permissive and restrictive configurations, keeping every entry changes the success rate by at most 2.9\%, and $\tau = 0.75$, which removes 39\% to 79\% of the entries kept at $\tau = 0.1$, by at most 10.9\%. The penalty has already pushed the unimportant entries close to zero, and an entry below 0.1 shifts its routing logit by less than $0.1 \alpha \sigma_l$, which rarely changes the selected experts. Such small entries still shift the gating weights of the selected experts for every token (Eq.~\ref{eq:steered_output}), and over the many tokens of a multi-turn conversation these shifts accumulate, while strong pruning removes experts that the broad defensive target needs.

\subsection{Defensive Capability Comparison}
\label{subsec:defensive_capability_comparison}

To position the defensive capabilities of \ourMethod within the current landscape of inference-time interventions, we compare our framework against SteerMoE~\cite{fayyaz2026steering}, a recent jailbreak defense method for MoE architectures that also does not require training or fine-tuning the target model. 

% Originally, SteerMoE is evaluated against several models, which include, among others, GPT-OSS-20B, Qwen3-30B-A3B-Instruct-2507, Mixtral-8x7B-Instruct-v0.1, and Phi-3.5-MoE-Instruct. To allow for a full comparison across all models, we also performed their defensive steering approach on DeepSeek-MoE-16B-Chat, Hunyuan-A13B-Instruct, and Qwen1.5-MoE-A2.7B-Chat. 

SteerMoE requires two sets of distinct behaviors on the same set of prompts to identify experts associated with a target behavior. The method analyzes the activations of the top-$k$ selected experts on the tokens following the final ``Assistant:'' to determine which experts correlate with safe refusals versus harmful compliance. For the unsafe responses, we utilize the same labeling setup and MHJ conversation histories used for the defensive configuration, as described in Section~\ref{subsec:datasets}. We then construct the corresponding safe responses by manually replacing all unsafe outputs with a direct refusal. To measure the success rate, we use the same Llama-Guard-3-8B evaluator employed in our other experiments and in the original SteerMoE implementation.

As shown in Table~\ref{tab:comparison}, \ourMethod consistently outperforms SteerMoE as a multi-turn jailbreak defense method across all evaluated models. \ourMethod achieves a defense success rate of 83.9\%, while SteerMoE achieves an average of 58.4\%, often only marginally above the unsteered model and below it on GPT-OSS-20B. Thus, while SteerMoE has demonstrated efficacy on single-turn jailbreak prompts, our results indicate that \ourMethod is substantially better for complex, multi-turn adversarial interactions.

We attribute this performance gap directly to how a method models expert behavior. SteerMoE finds safety experts using only the activations from the top-$k$ selected experts in the final model response. This approach suffers from three blind spots. First, discarding the preceding conversation history hides multi-turn jailbreak intentions, which are deliberately fragmented across several prompts. Second, relying on activation frequency discards the sequential nature of language processing and token routing, which carries information that a frequency count cannot represent. Section~\ref{subsec:mechanistic_baselines} isolates this factor: an LSTM that reads the tokens in random order finds masks whose defense success rate is 16.7\% and 31.5\% lower on Phi-3.5-MoE-Instruct and Qwen3-30B-A3B-Instruct-2507. Third, evaluating only the top-$k$ experts collapses the continuous routing logit distribution into a hard binary signal (selected versus unselected). This hard signal discards the latent safety information of the logits of unselected experts. If a safety expert consistently ranks just below the top-$k$ routing threshold (e.g., as the $k+1$ expert), SteerMoE remains blind to its existence. Consequently, it will not discover or steer towards all relevant safety experts. Conversely, \ourMethod finds safety experts by leveraging an LSTM surrogate model trained on the entire sequential conversation context and the full, continuous distribution of all expert logits. This allows our framework to capture both the temporal routing patterns across conversational tokens and the cross-layer dependencies that emerge during multi-turn conversations.

These observations highlight three advantages of \ourMethodNoSpace. Leveraging the complete conversation context to produce logits for analysis provides valuable information for finding safety-relevant experts. Modeling those logits as a sequence rather than as aggregate counts preserves where in a prompt each routing decision falls. Using the routing logits for analysis provides a more comprehensive mapping of safety-related behavior than top-$k$ expert selections alone.

\begin{table}[!t]
  \centering
  \caption{Success rates for multi-turn jailbreak defense. We compare \ourMethod against SteerMoE.}
  \label{tab:comparison}
  \setlength{\tabcolsep}{5.4pt}
  \begin{tabular}{l|ccc}
    \toprule
    \textbf{Model} & \textbf{Baseline} & \textbf{\ourMethodNoSpace} & \textbf{SteerMoE} \\
    \midrule
    DeepSeek-MoE-16B-Chat       & 52.8\% & 84.8\% & 59.6\% \\
    GPT-OSS-20B                 & 62.0\% & 88.0\% & 61.2\% \\
    Hunyuan-A13B-Instruct       & 48.8\% & 80.4\% & 51.2\% \\
    Mixtral-8x7B-Instruct-v0.1  & 47.6\% & 77.2\% & 66.8\% \\
    Phi-3.5-MoE-Instruct        & 57.6\% & 80.8\% & 59.2\% \\
    Qwen1.5-MoE-A2.7B-Chat      & 51.6\% & 87.2\% & 58.0\% \\
    Qwen3-30B-A3B-Instruct-2507 & 47.2\% & 89.2\% & 52.8\% \\
    \midrule
    \emph{Average} & \emph{52.5\%} & \emph{83.9\%} & \emph{58.4\%} \\
  \bottomrule
  \end{tabular}
\end{table}

\textbf{Activation versus Expert Steering.} The continuous shift of \ourMethod also matters. When we replace it with the rigid expert steering of SteerMoE and SafeX~\cite{lai2025safex}, which forces the selected experts into the top-$k$, the average defense success rate falls from 83.9\% to 69.0\% (Table~\ref{tab:activation_expert_steering}). Appendix~\ref{sec:activation_steering_versus_expert_steering} discusses this variant.

\subsection{Mechanistic Baselines}
\label{subsec:mechanistic_baselines}

\ourMethod rests on two claims. First, the increase in success rate over the unsteered model comes from which experts the mask steers. Second, the sequential modeling of behavior (Section~\ref{subsec:sequential_modeling_of_behavior}) is what lets the mask optimization find the experts that drive the target behavior. We test both with controls that each change one part of the pipeline and keep everything else fixed. We run them on Phi-3.5-MoE-Instruct and Qwen3-30B-A3B-Instruct-2507, the two models that we evaluate in all three configurations, and report each model separately. Table~\ref{tab:mechanistic_baselines} compares every control with the unsteered model and with \ourMethod on the same model, configuration, and test set.

\textbf{Specificity of the Experts.} If \ourMethod correctly identifies the experts that drive the target behavior, the same mask entries placed on other experts should fail to steer that behavior. Two controls rearrange the non-zero entries of the \ourMethod mask, keep their number and values, and apply them at the same amplitude $\alpha$. The \emph{Global Random Mask} permutes the entries across all layers and experts, which shows whether the identity of the steered experts matters or only the amount of intervention. The \emph{Layer-wise Permuted Mask} permutes them among the experts of their own layer. It keeps the depth at which the mask intervenes, so it shows whether the choice of expert within a layer matters. Each control is reported as the mean over three repetitions. In the permissive configuration, both controls fall below the unsteered model: the Layer-wise Permuted Mask reaches 57.6\% on Phi-3.5-MoE-Instruct and 32.5\% on Qwen3-30B-A3B-Instruct-2507, against 61.2\% and 33.2\% without steering and 93.0\% and 67.1\% with \ourMethodNoSpace, and the Global Random Mask falls further, to 52.1\% and 25.4\%. The defensive configuration shows the same pattern, with both controls below the unsteered rate on both models. In the restrictive configuration, the Global Random Mask stays at the unsteered rate, but the Layer-wise Permuted Mask keeps part of the effect, reaching 58.9\% on Phi-3.5-MoE-Instruct and 27.6\% on Qwen3-30B-A3B-Instruct-2507, against 34.3\% and 20.2\% without steering and 78.1\% and 84.3\% with \ourMethodNoSpace. For political neutrality, the depth at which the mask intervenes thus carries part of the effect, and the experts that \ourMethod selects within each layer carry the rest. Placed on other experts, the same entries at the same amplitude lower the success rate or recover only part of the effect, so the effect of \ourMethod is tied to which experts it steers, not to how much it perturbs the router.

\textbf{Sequential Modeling.} If the sequential modeling of Section~\ref{subsec:sequential_modeling_of_behavior} is what lets the mask optimization find the experts that drive the target behavior, surrogates that drop part of this modeling should find weaker masks, and the part that costs the most shows which property carries the effect. The surrogate can be replaced without changing mask creation or application, because the method only needs gradients with respect to the scaled routing logits of Eq.~\ref{eq:adaptive_scaling}. Three surrogates simplify the LSTM one step at a time. The \emph{Shuffled-order LSTM} keeps the architecture and permutes the tokens of every sequence at random during training and mask optimization, which removes the token order. \emph{BoE-MLP} averages the per-token representation $\mathbf{z}_t$ of Eq.~\ref{eq:concatenation} over the tokens and classifies it with one hidden layer of $H = 64$ units, the hidden size of the LSTM; the pooling also removes the combination of layers within each token. \emph{BoE-LR} classifies the same pooled vector with a linear head, which also removes the nonlinearity. Its mask amounts to the thresholded weights of a linear probe, the MoE analogue of probe-derived steering vectors in dense models~\cite{arditi2024refusal}. Each surrogate is trained on the profiling data of the LSTM, and $\lambda$ and $\alpha$ are selected again with the grid search and selection rule of Section~\ref{subsec:mask_settings}, with $\tau = 0.1$. Table~\ref{tab:mechanistic_baselines} gives the mean validation accuracy of each surrogate over the three repetitions as a subscript. Removing the token order costs the most, in every configuration and on both models: the Shuffled-order LSTM finds masks whose success rate is 16.6\% to 49.6\% lower than that of \ourMethodNoSpace, with the largest losses in the restrictive configuration (36.9\% and 49.6\% on Phi-3.5-MoE-Instruct and Qwen3-30B-A3B-Instruct-2507). The later steps cost less and not consistently: removing the combination of layers changes the success rate by $-$5.6\% to $+$8.8\%, and removing the nonlinearity lowers it further in five of the six cases, in which the linear probe of BoE-LR stays within 2.4\% of the unsteered model. The order of the routing logits therefore carries the effect. The surrogates differ less in validation accuracy than in the success of their masks: removing the order lowers the validation accuracy by only 1.7\% to 8.0\%, so the advantage of the LSTM lies in locating the experts that produce the behavior rather than in predicting the behavior. 

Together, the two kinds of controls show that the increase in success rate comes from which experts the mask steers and that the sequential modeling of behavior is what lets the mask optimization find these experts. The order of the routing logits, which carries this effect, is the information that the frequency count of SteerMoE discards (Section~\ref{subsec:defensive_capability_comparison}).

\begin{table*}[!t]
  \centering
  \caption{Mechanistic baselines for Phi-3.5-MoE-Instruct and Qwen3-30B-A3B-Instruct-2507: success rate (\%) on the test set, with the mean validation accuracy (\%) of the surrogates that produced the mask as a subscript. Each control's success rate is the mean over three repetitions. Every mask raises the side effect of its configuration by less than 5 percentage points.}
  \label{tab:mechanistic_baselines}
  \footnotesize
  \begin{tabular}{ll|cc|cc|ccc}
    \toprule
     & & & & \multicolumn{2}{c|}{\textbf{Mask controls}} & \multicolumn{3}{c}{\textbf{Surrogate controls}} \\
    \textbf{Configuration} & \textbf{Model} & \textbf{Baseline} & \textbf{\ourMethod} & \textbf{Global} & \textbf{Layer-wise} & \textbf{Shuffled} & \textbf{BoE-MLP} & \textbf{BoE-LR} \\
    \midrule
    \multirow{2}{*}{Defensive} & Phi-3.5-MoE-Instruct  & 57.6 & $80.8_{91.5}$ & 51.6 & 56.3 & $64.1_{88.4}$ & $62.4_{85.2}$ & $57.9_{80.6}$ \\
                               & Qwen3-30B-A3B-Instruct-2507    & 47.2 & $89.2_{94.5}$ & 41.1 & 45.7 & $57.7_{89.1}$ & $55.2_{87.8}$ & $48.5_{84.7}$ \\
    \midrule
    \multirow{2}{*}{Permissive} & Phi-3.5-MoE-Instruct  & 61.2 & $93.0_{92.0}$ & 52.1 & 57.6 & $76.4_{90.3}$ & $70.8_{84.1}$ & $73.5_{79.1}$ \\
                               & Qwen3-30B-A3B-Instruct-2507    & 33.2 & $67.1_{93.4}$ & 25.4 & 32.5 & $41.8_{85.4}$ & $44.3_{85.6}$ & $35.1_{83.5}$ \\
    \midrule
    \multirow{2}{*}{Restrictive} & Phi-3.5-MoE-Instruct  & 34.3 & $78.1_{87.2}$ & 33.8 & 58.9 & $41.2_{83.0}$ & $50.0_{81.6}$ & $33.8_{78.5}$ \\
                               & Qwen3-30B-A3B-Instruct-2507    & 20.2 & $84.3_{85.0}$ & 19.4 & 27.6 & $34.7_{81.4}$ & $38.4_{80.9}$ & $22.6_{80.0}$ \\
    \bottomrule
  \end{tabular}
\end{table*}

\section{Discussion}
\label{sec:discussion}
\textbf{Computational Cost and Overhead of \ourMethodNoSpace.} A primary advantage of \ourMethod over existing MoE defenses is its high computational efficiency. Our approach avoids modifying the weights of the MoE LLM, meaning the computational cost is strictly bound to the model size and standard inference, with zero requirement for costly full-parameter or LoRA fine-tuning. The only training overhead stems from the lightweight LSTM-based surrogate model. In our experiments, training the LSTM requires approximately five minutes on a single NVIDIA H100 GPU. Furthermore, the operational overhead at inference time is analytically strictly bounded. Because the steering mask $\hat{\mathbf{S}}$ is static and computed entirely offline, the online intervention consists solely of an $O(E)$ element-wise addition to the pre-routing logits, where $E$ is the total number of experts per layer (e.g., $E \le 128$ for all evaluated models). Compared to the dense multiply-accumulate operations executed during the standard attention and expert MLP computations of a forward pass, this isolated $O(E)$ vector addition introduces negligible measurable latency overhead, making \ourMethod highly practical for real-time, large-scale deployments.

\textbf{Limitations.} Despite its efficiency and effectiveness, \ourMethod has several limitations. First, our method relies on an LSTM surrogate model to approximate the complex cross-layer routing dynamics of the MoE architecture. While this approximation is effective in practice, inconsistent or highly non-linear routing behaviors in extraordinarily deep models may exceed the surrogate's ability to map the routing logits to behavior, potentially leading to less effective mask optimizations. Second, \ourMethod intervenes exclusively at the routing level. Because we do not alter the underlying expert weights, the defense assumes that the model possesses the inherent capacity to generate safe responses; if a model's experts are fundamentally poisoned or lack alignment data entirely, routing steering alone cannot synthesize safe behavior. Finally, the steering masks optimized in this work are static during inference. While they generalize well to unseen prompts within the target distribution, they may be less resilient to sophisticated, out-of-distribution, zero-day jailbreaks that aggressively shift the model's activation space.

\textbf{Future Work.} Our findings open several promising directions for future research. A natural extension is the development of dynamic, input-dependent steering masks. Rather than optimizing a static steering matrix, future iterations of \ourMethod could employ a classifier to predict the necessary steering mask on a per-prompt or per-token basis, dynamically adapting the model's safety to the immediate threat level, instead of predicting it before inference. Additionally, while this work focuses heavily on safety and alignment, the foundational framework of \ourMethod is task-agnostic. Future work should explore applying \ourMethod to other critical domains, such as on-the-fly domain adaptation (e.g., steering a general MoE to act as a specialized medical or legal expert), reducing hallucinations, or controlling personality traits in conversational agents. Finally, exploring the connection between routing steering and attention-head interventions could yield a unified framework for complete mechanistic control over sparse architectures.
\section{Related Work}
\label{sec:related_work}

Work on the internal mechanisms of refusal and jailbreaks in LLMs either finds linear directions in model activations that relate to safety-relevant features or identifies model components that contribute to refusal.

% A broad line of work explores the internal mechanisms behind refusal and jailbreaks for general Large Language Models. There is significant variation in the approaches used. Some works focus on finding linear directions in model activations related to safety-relevant features, while others aim to find specific model components that contribute to refusal.

Within the first line, representation engineering~\cite{zou2023representation} introduced the extraction of and intervention on concept directions in activation space. Arditi et al.~\cite{arditi2024refusal} showed that refusal is mediated by a single direction, along which interventions can jailbreak models. Wollschläger et al.~\cite{wollschlager2025the} found that multiple independent directions influence refusal, and Zhao et al.~\cite{zhao2026llms} demonstrated that models represent harmfulness separately from refusal. On the defense side, probes for harmful behavior are a common, computationally cheap strategy in practice~\cite{kramar2026building,cunningham2026constitutional}, and other methods identify safety-relevant subspaces and steer models to remain in certain regions~\cite{lee2025programming,zeng2025safesteeradaptivesubspacesteering,lu2026assistant}. However, these steering methods are designed for dense models and have no implementation for MoEs, which precludes a direct comparison.

% Within the features line of work, representation engineering~\cite{zou2023representation} introduced the broader framework of extracting and intervening on concept directions in activation space. Arditi et al.~\cite{arditi2024refusal} applied this lens to refusal specifically, showing that refusal is mediated by a single direction, and that interventions along this direction can effectively jailbreak models. In later work, this has been expanded into more effective ways to find relevant safety behaviors; Wollschläger et al.~\cite{wollschlager2025the} found that multiple independent directions influence refusal, and Zhao et al.~\cite{zhao2026llms} demonstrated that models separately represent harmfulness from refusal. On the defense side, the identification of these safety-relevant activation directions has led to several approaches. Probe-based approaches for identifying harmful behaviors are a common, computationally cheap strategy in practice~\cite{kramar2026building,cunningham2026constitutional}. Furthermore, methods that identify safety-relevant subspaces and steer models to remain in certain regions have been proposed~\cite{lee2025programming,zeng2025safesteeradaptivesubspacesteering,lu2026assistant}. However, approaches like SafeSteer~\cite{zeng2025safesteeradaptivesubspacesteering} are currently designed for dense models and lack implementations for MoEs, precluding direct comparison.

Other works identify architectural components behind safety behavior. Li et al.~\cite{li2025safety} identified safety layers that contribute significantly to refusal and can be used for targeted fine-tuning. Zhou et al.~\cite{zhou2025on} showed that ablating a single attention head can significantly degrade safety. At a finer granularity, small sets of ``safety neurons'' are used to restore safety on unsafe queries~\cite{chen2026towards} or as an objective to optimize attacks~\cite{wu2026neurostrike}. SafeSeek~\cite{yu2026safeseekuniversalattributionsafety} unifies these granularities by extracting computational subgraphs of weights, neurons, and attention heads. While these analyses are effective for dense models, the sparse activation of MoE models calls for analyzing a different component: expert routing.

% Alternatively, other works identify specific architectural components that contribute to relevant behaviors. Li et al.~\cite{li2025safety} identified safety layers contributing significantly to refusal, which can be used for targeted fine-tuning. Zhou et al.~\cite{zhou2025on} explored the role of attention heads, showing that ablating a single head can result in significant safety degradation. At a finer granularity, small subsets of neurons have been identified as ``safety neurons'', which are then used to restore safety on unsafe queries~\cite{chen2026towards}, or as an objective to optimize attacks~\cite{wu2026neurostrike}. These concepts are unified in SafeSeek~\cite{yu2026safeseekuniversalattributionsafety}, where the authors extract computational subgraphs across multiple architectural granularities, including individual weights, neurons, and attention heads. While these component-based analyses are highly effective for dense models, the sparse activation paradigm of Mixture-of-Experts necessitates analyzing a fundamentally different component: expert routing mechanisms.

Recent works study expert activations in MoE models to localize safety-relevant components, either to attack or to preserve alignment. GateBreaker~\cite{wu2025gatebreaker} is an inference-time attack that identifies safety experts, localizes safety structures within them, and disables them, compromising the safety alignment of the MoE model. On the defense side, SafeX~\cite{lai2025safex} shows that safety is concentrated in a small number of experts and fine-tunes these experts, an expensive operation that leaves routing-based safety shortcuts unexamined. SteerMoE~\cite{fayyaz2026steering} instead operates at inference time without fine-tuning and controls behaviors such as faithfulness by (de)activating experts. It isolates safety experts by the difference in activation rates between prompt sets with faithful and unfaithful responses, and at inference overrides the router by setting the logits of these experts to $\pm \infty$.%positive or negative infinity.

% Recent works have begun studying expert activations in MoE models to localize safety-relevant components. These works can be broadly categorized into attacks aimed at compromising alignment and defenses aimed at preserving it. GateBreaker~\cite{wu2025gatebreaker} introduces a three-stage attack framework deployed at inference time. The attack identifies safety experts and localizes safety structures within them in order to disable them, successfully compromising the safety alignment of the MoE model. Because GateBreaker is inherently an attack mechanism designed to break alignment, it cannot be evaluated as a defense baseline against protective steering methods without fundamental, non-trivial adaptations. On the MoE defense side, SafeX~\cite{lai2025safex} demonstrates that safety is concentrated in a small number of experts and proposes fine-tuning these experts to improve safety, requiring a relatively expensive operation that leaves routing-based safety shortcuts unexamined. In contrast, SteerMoE~\cite{fayyaz2026steering} operates at inference time without fine-tuning, aiming to control behaviors like faithfulness by selectively (de)activating experts. It relies on a frequency-based analysis of activation rate differences between prompt sets representing faithful and unfaithful responses to isolate safety experts. During inference, SteerMoE overrides the router by explicitly setting the logits of certain experts to positive or negative infinity.

\section{Conclusion}
\label{sec:conclusion}

In this paper, we introduced \ourMethodNoSpace, a lightweight framework that selectively configures the safety-related behavior of MoE architectures without training or fine-tuning the target model. By utilizing an LSTM-based surrogate model trained on continuous routing logits, our approach successfully maps complex routing dependencies to downstream behavioral experts. Through the application of static, sparse steering masks to the routing logits, \ourMethod biases expert selection to enhance or suppress specific behaviors without the prohibitive costs of retraining or inference delays.

Our evaluation across seven diverse open-source MoE models demonstrates the framework's high efficacy and versatility, while preserving model utility and producing only limited side effects. \ourMethod increased the average defense success rate from 52.5\% to 83.9\%, increased the average generation success rate for adult-content requests from 52.6\% to 82.0\%, and raised the share of political questions on which the models take no position from 31.8\% to 75.4\%. The results further show that the learned defensive mask retains substantial effectiveness against a defense-aware adaptive attacker. Together, these findings show that \ourMethod provides a useful control surface for lightweight, deployment-time modification of safety-related model behavior.

\bibliographystyle{IEEEtran}
\bibliography{bibliography}

@misc{zou2023universal,
      title={Universal and Transferable Adversarial Attacks on Aligned Language Models}, 
      author={Andy Zou and Zifan Wang and J. Zico Kolter and Matt Fredrikson},
      year={2023},
      eprint={2307.15043},
      archivePrefix={arXiv},
      primaryClass={cs.CL}
}

@misc{li2024llmdefensesrobustmultiturn,
      title={{LLM} Defenses Are Not Robust to Multi-Turn Human Jailbreaks Yet}, 
      author={Nathaniel Li and Ziwen Han and Ian Steneker and Willow Primack and Riley Goodside and Hugh Zhang and Zifan Wang and Cristina Menghini and Summer Yue},
      year={2024},
      eprint={2408.15221},
      archivePrefix={arXiv},
      primaryClass={cs.LG},
      url={https://arxiv.org/abs/2408.15221}, 
}

@misc{song2026multibreak,
      title={{MultiBreak}: A Scalable and Diverse Multi-turn Jailbreak Benchmark for Evaluating {LLM} Safety}, 
      author={Jialin Song and Xiaodong Liu and Weiwei Yang and Wuyang Chen and Mingqian Feng and Xuekai Zhu and Jianfeng Gao},
      year={2026},
      eprint={2605.01687},
      archivePrefix={arXiv},
      primaryClass={cs.CL},
      url={https://arxiv.org/abs/2605.01687}, 
}

@misc{openerotica2024eroticaanalysis,
      author={{OpenErotica}},
      title={erotica-analysis: A Dataset for Erotica Literature Analysis},
      year={2024},
      publisher={Hugging Face},
      howpublished={\url{https://huggingface.co/datasets/openerotica/erotica-analysis}},
      note={Accessed: April 5, 2026}
}

@misc{yuan2025naturalreasoningreasoningwild28m,
      title={{NaturalReasoning}: Reasoning in the Wild with {2.8M} Challenging Questions}, 
      author={Weizhe Yuan and Jane Yu and Song Jiang and Karthik Padthe and Yang Li and Dong Wang and Ilia Kulikov and Kyunghyun Cho and Yuandong Tian and Jason E Weston and Xian Li},
      year={2025},
      eprint={2502.13124},
      archivePrefix={arXiv},
      primaryClass={cs.CL},
      url={https://arxiv.org/abs/2502.13124}, 
}

@misc{kingma2017adam,
      title={Adam: A Method for Stochastic Optimization}, 
      author={Diederik P. Kingma and Jimmy Ba},
      year={2017},
      eprint={1412.6980},
      archivePrefix={arXiv},
      primaryClass={cs.LG},
      url={https://arxiv.org/abs/1412.6980}, 
}

@misc{grattafiori2024llama3herdmodels,
      title={The {Llama} 3 Herd of Models}, 
      author={Aaron Grattafiori and Abhimanyu Dubey and Abhinav Jauhri and Abhinav Pandey and Abhishek Kadian and Ahmad Al-Dahle and Aiesha Letman and Akhil Mathur and Alan Schelten and Alex Vaughan and others},
      year={2024},
      eprint={2407.21783},
      archivePrefix={arXiv},
      primaryClass={cs.AI},
      url={https://arxiv.org/abs/2407.21783}, 
}

@inproceedings{fayyaz2026steering,
      title={Steering {MoE} {LLM}s via Expert (De)Activation},
      author={Mohsen Fayyaz and Ali Modarressi and Hanieh Deilamsalehy and Franck Dernoncourt and Ryan A. Rossi and Trung Bui and Hinrich Schuetze and Nanyun Peng},
      booktitle={The Fourteenth International Conference on Learning Representations},
      year={2026},
      url={https://openreview.net/forum?id=v5Yl9V8rJs}
}

@misc{zeng2025safesteeradaptivesubspacesteering,
      title={{SafeSteer}: Adaptive Subspace Steering for Efficient Jailbreak Defense in Vision-Language Models}, 
      author={Xiyu Zeng and Siyuan Liang and Liming Lu and Haotian Zhu and Enguang Liu and Jisheng Dang and Yongbin Zhou and Shuchao Pang},
      year={2025},
      eprint={2509.21400},
      archivePrefix={arXiv},
      primaryClass={cs.CR},
      url={https://arxiv.org/abs/2509.21400}, 
}

@misc{yu2026safeseekuniversalattributionsafety,
      title={{SafeSeek}: Universal Attribution of Safety Circuits in Language Models}, 
      author={Miao Yu and Siyuan Fu and Moayad Aloqaily and Zhenhong Zhou and Safa Otoum and Xing fan and Kun Wang and Yufei Guo and Qingsong Wen},
      year={2026},
      eprint={2603.23268},
      archivePrefix={arXiv},
      primaryClass={cs.LG},
      url={https://arxiv.org/abs/2603.23268}, 
}

@misc{dai2024deepseekmoe,
      title={{DeepSeekMoE}: Towards Ultimate Expert Specialization in Mixture-of-Experts Language Models}, 
      author={Damai Dai and Chengqi Deng and Chenggang Zhao and R. X. Xu and Huazuo Gao and Deli Chen and Jiashi Li and Wangding Zeng and Xingkai Yu and Y. Wu and others},
      year={2024},
      eprint={2401.06066},
      archivePrefix={arXiv},
      primaryClass={cs.CL},
      url={https://arxiv.org/abs/2401.06066}
}

@misc{openai2025gptossmoe,
      title={{Introducing GPT-OSS}},
      url={https://openai.com/index/introducing-gpt-oss/},
      author={{OpenAI}},
      month={August},
      year={2025}
}

@misc{openai2025adult,
      title={{OpenAI will allow verified adults to use ChatGPT to generate erotic content}},
      url={https://www.theguardian.com/technology/2025/oct/14/openai-chatgpt-adult-erotic-content},
      author={{Guardian}},
      month={October},
      year={2025}
}

@misc{anthropic2026fable,
      title={{Redeploying Fable 5}},
      url={https://www.anthropic.com/news/redeploying-fable-5},
      author={{Anthropic}},
      month={June},
      year={2026}
}

@misc{tencent2025hunyuanmoe,
      title={{Hunyuan-A13B Technical Report}},
      url={https://github.com/Tencent-Hunyuan/Hunyuan-A13B/blob/main/report/Hunyuan_A13B_Technical_Report.pdf},
      author={{Hunyuan Team Tencent}},
      month={June},
      year={2025},
}

@misc{jiang2024mixtralmoe,
      title={Mixtral of Experts}, 
      author={Albert Q. Jiang and Alexandre Sablayrolles and Antoine Roux and Arthur Mensch and Blanche Savary and Chris Bamford and Devendra Singh Chaplot and Diego de las Casas and Emma Bou Hanna and Florian Bressand and others},
      year={2024},
      eprint={2401.04088},
      archivePrefix={arXiv},
      primaryClass={cs.LG},
      url={https://arxiv.org/abs/2401.04088}, 
}

@misc{abdin2024phi3moe,
      title={Phi-3 Technical Report: A Highly Capable Language Model Locally on Your Phone}, 
      author={Marah Abdin and Jyoti Aneja and Hany Awadalla and Ahmed Awadallah and Ammar Ahmad Awan and Nguyen Bach and Amit Bahree and Arash Bakhtiari and Jianmin Bao and Harkirat Behl and others},
      year={2024},
      eprint={2404.14219},
      archivePrefix={arXiv},
      primaryClass={cs.CL},
      url={https://arxiv.org/abs/2404.14219}, 
}

@misc{alibaba2024qwen15moe,
      title={{Qwen1.5-MoE: Matching 7B Model Performance with 1/3 Activated Parameters}},
      url={https://qwenlm.github.io/blog/qwen-moe/},
      author={{Qwen Team}},
      month={February},
      year={2024}
}

@misc{yang2025qwen3moe,
      title={Qwen3 Technical Report}, 
      author={An Yang and Anfeng Li and Baosong Yang and Beichen Zhang and Binyuan Hui and Bo Zheng and Bowen Yu and Chang Gao and Chengen Huang and Chenxu Lv and others},
      year={2025},
      eprint={2505.09388},
      archivePrefix={arXiv},
      primaryClass={cs.CL},
      url={https://arxiv.org/abs/2505.09388}, 
}

@inproceedings{lai2025safex,
      title={{SAFE}x: Analyzing Vulnerabilities of {MoE}-Based {LLM}s via Stable Safety-critical Expert Identification},
      author={ZhengLin Lai and Mengyao Liao and Bingzhe Wu and Dong Xu and Zebin Zhao and Zhihang Yuan and Chao Fan and Jianqiang Li},
      booktitle={The Thirty-ninth Annual Conference on Neural Information Processing Systems},
      year={2025},
      url={https://openreview.net/forum?id=VwsXmcMyg5}
}

@misc{shazeer2017outrageouslylargeneuralnetworks,
      title={Outrageously Large Neural Networks: The Sparsely-Gated Mixture-of-Experts Layer}, 
      author={Noam Shazeer and Azalia Mirhoseini and Krzysztof Maziarz and Andy Davis and Quoc Le and Geoffrey Hinton and Jeff Dean},
      year={2017},
      eprint={1701.06538},
      archivePrefix={arXiv},
      primaryClass={cs.LG},
      url={https://arxiv.org/abs/1701.06538}, 
}

@misc{wu2025gatebreaker,
      title={{GateBreaker}: Gate-Guided Attacks on Mixture-of-Expert {LLMs}}, 
      author={Lichao Wu and Sasha Behrouzi and Mohamadreza Rostami and Stjepan Picek and Ahmad-Reza Sadeghi},
      year={2025},
      eprint={2512.21008},
      archivePrefix={arXiv},
      primaryClass={cs.CR},
      url={https://arxiv.org/abs/2512.21008}, 
}

@misc{lintelo2026largelanguagelobotomy,
      title={Large Language Lobotomy: Jailbreaking Mixture-of-Experts via Expert Silencing}, 
      author={Jona te Lintelo and Lichao Wu and Stjepan Picek},
      year={2026},
      eprint={2602.08741},
      archivePrefix={arXiv},
      primaryClass={cs.CR},
      url={https://arxiv.org/abs/2602.08741}, 
}

@inproceedings{hendrycks2021measuring,
      title={Measuring Massive Multitask Language Understanding},
      author={Dan Hendrycks and Collin Burns and Steven Basart and Andy Zou and Mantas Mazeika and Dawn Song and Jacob Steinhardt},
      booktitle={International Conference on Learning Representations},
      year={2021},
      url={https://openreview.net/forum?id=d7KBjmI3GmQ}
}

@misc{cobbe2021gsm8k,
      title={Training Verifiers to Solve Math Word Problems}, 
      author={Karl Cobbe and Vineet Kosaraju and Mohammad Bavarian and Mark Chen and Heewoo Jun and Lukasz Kaiser and Matthias Plappert and Jerry Tworek and Jacob Hilton and Reiichiro Nakano and others},
      year={2021},
      eprint={2110.14168},
      archivePrefix={arXiv},
      primaryClass={cs.LG},
      url={https://arxiv.org/abs/2110.14168}, 
}

@inproceedings{brown2020fewshotlearners,
      author={Brown, Tom and Mann, Benjamin and Ryder, Nick and Subbiah, Melanie and Kaplan, Jared D and Dhariwal, Prafulla and Neelakantan, Arvind and Shyam, Pranav and Sastry, Girish and Askell, Amanda and others},
      booktitle={Advances in Neural Information Processing Systems},
      editor={H. Larochelle and M. Ranzato and R. Hadsell and M.F. Balcan and H. Lin},
      pages={1877--1901},
      publisher={Curran Associates, Inc.},
      title={Language Models are Few-Shot Learners},
      url={https://proceedings.neurips.cc/paper_files/paper/2020/file/1457c0d6bfcb4967418bfb8ac142f64a-Paper.pdf},
      volume={33},
      year={2020}
}

@misc{chen2021evaluatinglargelanguagemodels,
      title={Evaluating Large Language Models Trained on Code}, 
      author={Mark Chen and Jerry Tworek and Heewoo Jun and Qiming Yuan and Henrique Ponde de Oliveira Pinto and Jared Kaplan and Harri Edwards and Yuri Burda and Nicholas Joseph and Greg Brockman and others},
      year={2021},
      eprint={2107.03374},
      archivePrefix={arXiv},
      primaryClass={cs.LG},
      url={https://arxiv.org/abs/2107.03374}, 
}

@inproceedings{kojima2022oneshotreasoners,
      author={Kojima, Takeshi and Gu, Shixiang (Shane) and Reid, Machel and Matsuo, Yutaka and Iwasawa, Yusuke},
      booktitle={Advances in Neural Information Processing Systems},
      editor={S. Koyejo and S. Mohamed and A. Agarwal and D. Belgrave and K. Cho and A. Oh},
      pages={22199--22213},
      publisher={Curran Associates, Inc.},
      title={Large Language Models are Zero-Shot Reasoners},
      url={https://proceedings.neurips.cc/paper_files/paper/2022/file/8bb0d291acd4acf06ef112099c16f326-Paper-Conference.pdf},
      volume={35},
      year={2022}
}

@article{wei2022emergent,
      title={Emergent Abilities of Large Language Models},
      author={Jason Wei and Yi Tay and Rishi Bommasani and Colin Raffel and Barret Zoph and Sebastian Borgeaud and Dani Yogatama and Maarten Bosma and Denny Zhou and Donald Metzler and Ed H. Chi and Tatsunori Hashimoto and Oriol Vinyals and Percy Liang and Jeff Dean and William Fedus},
      journal={Transactions on Machine Learning Research},
      issn={2835-8856},
      year={2022},
      url={https://openreview.net/forum?id=yzkSU5zdwD},
}

@article{hu2022lora,
      title={Lora: Low-rank adaptation of large language models.},
      author={Hu, Edward J and Shen, Yelong and Wallis, Phillip and Allen-Zhu, Zeyuan and Li, Yuanzhi and Wang, Shean and Wang, Liang and Chen, Weizhu and others},
      journal={Iclr},
      volume={1},
      number={2},
      pages={3},
      year={2022}
}

@misc{lialin2024scalingscaleupguide,
      title={Scaling Down to Scale Up: A Guide to Parameter-Efficient Fine-Tuning}, 
      author={Vladislav Lialin and Vijeta Deshpande and Xiaowei Yao and Anna Rumshisky},
      year={2024},
      eprint={2303.15647},
      archivePrefix={arXiv},
      primaryClass={cs.CL},
      url={https://arxiv.org/abs/2303.15647}, 
}

@article{patterson2022carbon,
      author={Patterson, David and Gonzalez, Joseph and Hölzle, Urs and Le, Quoc and Liang, Chen and Munguia, Lluis-Miquel and Rothchild, Daniel and So, David R. and Texier, Maud and Dean, Jeff},
      journal={Computer}, 
      title={The Carbon Footprint of Machine Learning Training Will Plateau, Then Shrink}, 
      year={2022},
      volume={55},
      number={7},
      pages={18-28},
      doi={10.1109/MC.2022.3148714}
}

@article{elhage2021mathematical,
      title={A Mathematical Framework for Transformer Circuits},
      author={Elhage, Nelson and Nanda, Neel and Olsson, Catherine and Henighan, Tom and Joseph, Nicholas and Mann, Ben and Askell, Amanda and Bai, Yuntao and Chen, Anna and Conerly, Tom and DasSarma, Nova and Drain, Dawn and Ganguli, Deep and Hatfield-Dodds, Zac and Hernandez, Danny and Jones, Andy and Kernion, Jackson and Lovitt, Liane and Ndousse, Kamal and Amodei, Dario and Brown, Tom and Clark, Jack and Kaplan, Jared and McCandlish, Sam and Olah, Chris},
      year={2021},
      journal={Transformer Circuits Thread},
      note={https://transformer-circuits.pub/2021/framework/index.html}
}

@article{fedus2022switchtransformers,
      author={Fedus, William and Zoph, Barret and Shazeer, Noam},
      title={Switch transformers: scaling to trillion parameter models with simple and efficient sparsity},
      year={2022},
      issue_date={January 2022},
      publisher={JMLR.org},
      volume={23},
      number={1},
      issn={1532-4435},
      journal={J. Mach. Learn. Res.},
      month=jan,
      articleno={120},
      numpages={39}
}

@inproceedings{
      cunningham2026constitutional,
      title={Constitutional Classifiers++: Efficient Production-Grade Defenses against Universal Jailbreaks},
      author={Hoagy Cunningham and Jerry Wei and Zihan Wang and Andrew Persic and Alwin Peng and Jordan Abderrachid and Raj Agarwal and Bobby Chen and Andy Dau and Alek Dimitriev and others},
      booktitle={The Fourteenth International Conference on Learning Representations},
      year={2026},
      url={https://openreview.net/forum?id=eNvsH5Ye2V}
}

@article{kramar2026building,
      title={Building Production-Ready Probes For {Gemini}},
      author={Kram{\'a}r, J{\'a}nos and Engels, Joshua and Wang, Zheng and Chughtai, Bilal and Shah, Rohin and Nanda, Neel and Conmy, Arthur},
      journal={arXiv preprint arXiv:2601.11516},
      year={2026}
}

@inproceedings{park2024the,
      title={The Linear Representation Hypothesis and the Geometry of Large Language Models},
      author={Kiho Park and Yo Joong Choe and Victor Veitch},
      booktitle={Forty-first International Conference on Machine Learning},
      year={2024},
      url={https://openreview.net/forum?id=UGpGkLzwpP}
}

@article{zou2023representation,
      title={Representation engineering: A top-down approach to {AI} transparency},
      author={Zou, Andy and Phan, Long and Chen, Sarah and Campbell, James and Guo, Phillip and Ren, Richard and Pan, Alexander and Yin, Xuwang and Mazeika, Mantas and Dombrowski, Ann-Kathrin and others},
      journal={arXiv preprint arXiv:2310.01405},
      year={2023}
}

@inproceedings{zhao2026llms,
      title={{LLM}s Encode Harmfulness and Refusal Separately},
      author={Jiachen Zhao and Jing Huang and Zhengxuan Wu and David Bau and Weiyan Shi},
      booktitle={The Thirty-ninth Annual Conference on Neural Information Processing Systems},
      year={2026},
      url={https://openreview.net/forum?id=zLkpt30ngy}
}

@inproceedings{lee2025programming,
      title={Programming Refusal with Conditional Activation Steering},
      author={Bruce W. Lee and Inkit Padhi and Karthikeyan Natesan Ramamurthy and Erik Miehling and Pierre Dognin and Manish Nagireddy and Amit Dhurandhar},
      booktitle={The Thirteenth International Conference on Learning Representations},
      year={2025},
      url={https://openreview.net/forum?id=Oi47wc10sm}
}

@inproceedings{li2025safety,
      title={Safety Layers in Aligned Large Language Models: The Key to {LLM} Security},
      author={Shen Li and Liuyi Yao and Lan Zhang and Yaliang Li},
      booktitle={The Thirteenth International Conference on Learning Representations},
      year={2025},
      url={https://openreview.net/forum?id=kUH1yPMAn7}
}

@inproceedings{zhou2025on,
      title={On the Role of Attention Heads in Large Language Model Safety},
      author={Zhenhong Zhou and Haiyang Yu and Xinghua Zhang and Rongwu Xu and Fei Huang and Kun Wang and Yang Liu and Junfeng Fang and Yongbin Li},
      booktitle={The Thirteenth International Conference on Learning Representations},
      year={2025},
      url={https://openreview.net/forum?id=h0Ak8A5yqw}
}

@article{wu2026neurostrike,
      title={{NeuroStrike}: Neuron-Level Attacks on Aligned {LLMs}},
      author={Wu, Lichao and Behrouzi, Sasha and Rostami, Mohamadreza and Thang, Maximilian and Picek, Stjepan and Sadeghi, Ahmad-Reza},
      journal={Network and Distributed System Security (NDSS) Symposium},
      year={2026}
}

@inproceedings{chen2026towards,
      title={Towards Understanding Safety Alignment: A Mechanistic Perspective from Safety Neurons},
      author={Jianhui Chen and Xiaozhi Wang and Zijun Yao and Yushi Bai and Lei Hou and Juanzi Li},
      booktitle={The Thirty-ninth Annual Conference on Neural Information Processing Systems},
      year={2026},
      url={https://openreview.net/forum?id=AAXMcAyNF6}
}

@inproceedings{wollschlager2025the,
      title={The Geometry of Refusal in Large Language Models: Concept Cones and Representational Independence},
      author={Tom Wollschl{\"a}ger and Jannes Elstner and Simon Geisler and Vincent Cohen-Addad and Stephan G{\"u}nnemann and Johannes Gasteiger},
      booktitle={Forty-second International Conference on Machine Learning},
      year={2025},
      url={https://openreview.net/forum?id=80IwJqlXs8}
}

@article{bricken2023monosemanticity,
      title={Towards Monosemanticity: Decomposing Language Models With Dictionary Learning},
      author={Bricken, Trenton and Templeton, Adly and Batson, Joshua and Chen, Brian and Jermyn, Adam and Conerly, Tom and Turner, Nick and Anil, Cem and Denison, Carson and Askell, Amanda and others},
      year={2023},
      journal={Transformer Circuits Thread},
      note={https://transformer-circuits.pub/2023/monosemantic-features/index.html}
}

@inproceedings{hua2026steering,
      title={Steering Evaluation-Aware Language Models To Act Like They Are Deployed},
      author={Tim Tian Hua and Andrew Qin and Samuel Marks and Neel Nanda},
      booktitle={The Fourteenth International Conference on Learning Representations},
      year={2026},
      url={https://openreview.net/forum?id=1TdRdf0fkw}
}

@article{arditi2024refusal,
      title={Refusal in language models is mediated by a single direction},
      author={Arditi, Andy and Obeso, Oscar and Syed, Aaquib and Paleka, Daniel and Panickssery, Nina and Gurnee, Wes and Nanda, Neel},
      journal={Advances in Neural Information Processing Systems},
      volume={37},
      pages={136037--136083},
      year={2024}
}

@article{chen2025persona,
      title={Persona vectors: Monitoring and controlling character traits in language models\allowbreak},
      author={Chen, Runjin and Arditi, Andy and Sleight, Henry and Evans, Owain and Lindsey, Jack},
      journal={arXiv preprint arXiv:2507.21509},
      year={2025}
}

@article{lu2026assistant,
      title={The assistant axis: Situating and stabilizing the default persona of language models},
      author={Lu, Christina and Gallagher, Jack and Michala, Jonathan and Fish, Kyle and Lindsey, Jack},
      journal={arXiv preprint arXiv:2601.10387},
      year={2026}
}

@article{wei2023jailbroken,
      title={Jailbroken: How does {LLM} safety training fail?},
      author={Wei, Alexander and Haghtalab, Nika and Steinhardt, Jacob},
      journal={Advances in Neural Information Processing Systems},
      volume={36},
      pages={80079--80110},
      year={2023}
}

@article{niu2024jailbreaking,
      title={Jailbreaking attack against multimodal large language model},
      author={Niu, Zhenxing and Ren, Haodong and Gao, Xinbo and Hua, Gang and Jin, Rong},
      journal={arXiv preprint arXiv:2402.02309},
      year={2024}
}

@inproceedings{shen2024anything,
      title={{{``Do Anything Now''}: Characterizing and Evaluating In-The-Wild Jailbreak Prompts on Large Language Models}},
      author={Shen, Xinyue and Chen, Zeyuan and Backes, Michael and Shen, Yun and Zhang, Yang},
      booktitle={Proceedings of the 2024 on ACM SIGSAC Conference on Computer and Communications Security},
      pages={1671--1685},
      year={2024}
}

@misc{gao2024evalharness,
      author={Gao, Leo and Tow, Jonathan and Abbasi, Baber and Biderman, Stella and Black, Sid and DiPofi, Anthony and Foster, Charles and Golding, Laurence and Hsu, Jeffrey and Le Noac'h, Alain and others},
      title={The Language Model Evaluation Harness},
      month=07,
      year=2024,
      publisher={Zenodo},
      version={v0.4.3},
      doi={10.5281/zenodo.12608602},
      url={https://zenodo.org/records/12608602}
}

@inproceedings{souly2024strongreject,
      author={Souly, Alexandra and Lu, Qingyuan and Bowen, Dillon and Trinh, Tu and Hsieh, Elvis and Pandey, Sana and Abbeel, Pieter and Svegliato, Justin and Emmons, Scott and Watkins, Olivia and Toyer, Sam},
      booktitle={Advances in Neural Information Processing Systems},
      doi={10.52202/079017-3984},
      editor={A. Globerson and L. Mackey and D. Belgrave and A. Fan and U. Paquet and J. Tomczak and C. Zhang},
      pages={125416--125440},
      publisher={Curran Associates, Inc.},
      title={A {StrongREJECT} for Empty Jailbreaks},
      url={https://proceedings.neurips.cc/paper_files/paper/2024/file/e2e06adf560b0706d3b1ddfca9f29756-Paper-Datasets_and_Benchmarks_Track.pdf},
      volume={37},
      year={2024}
}

@misc{zhou2023instruction,
      title={Instruction-Following Evaluation for Large Language Models}, 
      author={Jeffrey Zhou and Tianjian Lu and Swaroop Mishra and Siddhartha Brahma and Sujoy Basu and Yi Luan and Denny Zhou and Le Hou},
      year={2023},
      eprint={2311.07911},
      archivePrefix={arXiv},
      primaryClass={cs.CL},
      url={https://arxiv.org/abs/2311.07911}, 
}

@inproceedings {russinovich2024crescendo,
      author={Mark Russinovich and Ahmed Salem and Ronen Eldan},
      title={Great, Now Write an Article About That: The {Crescendo} {Multi-Turn} {LLM} Jailbreak Attack},
      booktitle={34th USENIX Security Symposium (USENIX Security 25)},
      year={2025},
      isbn={978-1-939133-52-6},
      address={Seattle, WA},
      pages={2421--2440},
      url={https://www.usenix.org/conference/usenixsecurity25/presentation/russinovich},
      publisher={USENIX Association},
      month=aug
}

@inproceedings{han2024wildguard,
author = {Han, Seungju and Rao, Kavel and Ettinger, Allyson and Jiang, Liwei and Lin, Bill Yuchen and Lambert, Nathan and Choi, Yejin and Dziri, Nouha},
title = {{WildGuard}: open one-stop moderation tools for safety risks, jailbreaks, and refusals of {LLMs}},
year = {2024},
isbn = {9798331314385},
publisher = {Curran Associates Inc.},
address = {Red Hook, NY, USA},
booktitle = {Proceedings of the 38th International Conference on Neural Information Processing Systems},
articleno = {261},
numpages = {39},
location = {Vancouver, BC, Canada},
series = {NIPS '24}
}

@misc{qwen2025qwen25,
      title={Qwen2.5 Technical Report}, 
      author={Qwen and An Yang and Baosong Yang and Beichen Zhang and Binyuan Hui and Bo Zheng and Bowen Yu and Chengyuan Li and Dayiheng Liu and Fei Huang and Haoran Wei and Huan Lin and Jian Yang and Jianhong Tu and Jianwei Zhang and Jianxin Yang and Jiaxi Yang and Jingren Zhou and Junyang Lin and Kai Dang and Keming Lu and Keqin Bao and Kexin Yang and Le Yu and Mei Li and Mingfeng Xue and Pei Zhang and Qin Zhu and Rui Men and Runji Lin and Tianhao Li and Tianyi Tang and Tingyu Xia and Xingzhang Ren and Xuancheng Ren and Yang Fan and Yang Su and Yichang Zhang and Yu Wan and Yuqiong Liu and Zeyu Cui and Zhenru Zhang and Zihan Qiu},
      year={2025},
      eprint={2412.15115},
      archivePrefix={arXiv},
      primaryClass={cs.CL},
      url={https://arxiv.org/abs/2412.15115}, 
}

@article{zhang2026healthorsc,
  title   = {Health-{ORSC}-Bench: A Benchmark for Measuring Over-Refusal and Safety Completion in Health Context},
  author  = {Zhang, Zhihao and Huang, Liting and Wu, Guanghao and Nakov, Preslav and Ji, Heng and Naseem, Usman},
  journal = {arXiv preprint arXiv:2601.17642},
  year    = {2026}
}

@misc{durmus2023towards,
      title={Towards Measuring the Representation of Subjective Global Opinions in Language Models}, 
      author={Esin Durmus and Karina Nyugen and Thomas I. Liao and Nicholas Schiefer and Amanda Askell and Anton Bakhtin and Carol Chen and Zac Hatfield-Dodds and Danny Hernandez and Nicholas Joseph and Liane Lovitt and Sam McCandlish and Orowa Sikder and Alex Tamkin and Janel Thamkul and Jared Kaplan and Jack Clark and Deep Ganguli},
      year={2023},
      eprint={2306.16388},
      archivePrefix={arXiv},
      primaryClass={cs.CL}
}

@inproceedings{tramer2020adaptive,
 author = {Tramer, Florian and Carlini, Nicholas and Brendel, Wieland and Madry, Aleksander},
 booktitle = {Advances in Neural Information Processing Systems},
 editor = {H. Larochelle and M. Ranzato and R. Hadsell and M.F. Balcan and H. Lin},
 pages = {1633--1645},
 publisher = {Curran Associates, Inc.},
 title = {On Adaptive Attacks to Adversarial Example Defenses},
 url = {https://proceedings.neurips.cc/paper_files/paper/2020/file/11f38f8ecd71867b42433548d1078e38-Paper.pdf},
 volume = {33},
 year = {2020}
}

@inproceedings{
kim2026safemoe,
title={{SafeMoE}: Safe Fine-Tuning for {MoE} {LLM}s by Aligning Harmful Input Routing},
author={Jaehan Kim and Minkyoo Song and Seungwon Shin and Sooel Son},
booktitle={The Fourteenth International Conference on Learning Representations},
year={2026},
url={https://openreview.net/forum?id=W1x9AzkSnU}
}

@article{hochreiter1997lstm,
  author  = {Hochreiter, Sepp and Schmidhuber, J{\"u}rgen},
  title   = {Long Short-Term Memory},
  journal = {Neural Computation},
  volume  = {9},
  number  = {8},
  pages   = {1735--1780},
  year    = {1997},
  doi     = {10.1162/neco.1997.9.8.1735}
}

\appendices

\section{Target Model Architectures and Hardware}
\label{app:architectures}

\begin{table*}[!t]
  \centering
  \caption{Architecture specifications of the targeted MoE LLMs.}
  \label{tab:model_specifications}
  \begin{tabular}{l|ccccccc}
    \toprule
    \textbf{Model} & \textbf{Sparse Experts} & \textbf{Shared Experts} & \textbf{MoE Layers} & \textbf{Top-$k$} & \textbf{Active/Total Params} & \textbf{Developer} \\
    \midrule
    DeepSeek-MoE-16B-Chat       & 64  & 2   & 27 & 6 & 2.7B / 16.4B  & DeepSeek  \\
    GPT-OSS-20B                 & 32  & N/A & 24 & 4 & 3.6B / 21B    & OpenAI    \\
    Hunyuan-A13B-Instruct       & 64  & 1   & 32 & 8 & 13B / 80.4B   & Tencent   \\
    Mixtral-8x7B-Instruct-v0.1  & 8   & N/A & 32 & 2 & 12.9B / 46.7B & Mistral   \\
    Phi-3.5-MoE-Instruct        & 16  & N/A & 32 & 2 & 6.6B / 41.9B  & Microsoft \\
    Qwen1.5-MoE-A2.7B-Chat      & 60  & 4   & 24 & 4 & 2.7B / 14.3B  & Alibaba   \\
    Qwen3-30B-A3B-Instruct-2507 & 128 & N/A & 48 & 8 & 3.3B / 30.5B  & Alibaba   \\
    \bottomrule
  \end{tabular}
\end{table*}

Table~\ref{tab:model_specifications} lists the architectural properties of the seven target models. We chose them to span the design space of current open MoE LLMs along every axis that could affect how safety behavior is distributed: the number of experts per layer, the number of experts selected per token, the total parameter count, and the architectural variant.

% The MoE Arch. column distinguishes these variants. \emph{Sparse} models route every token to their top-$k$ experts and have no other expert computation. \emph{Shared} models additionally pass every token through one or more shared experts that are always active. \emph{Grouped} denotes the Mixture of Grouped Experts design of Pangu-Pro, which partitions the routed experts into equal groups and selects the same number of experts from each group for every token. \ourMethod acts only on routed experts, since silencing operates on the router's logits and shared experts bypass the router. Shared experts are therefore excluded from every count in the table and are never silenced.

All implementations and evaluations in this work were conducted using CUDA-enabled GPUs for optimal runtimes. To fit the memory footprint of the target models and inference, we used two NVIDIA GH200 120 GB Grace Hopper Superchips paired with NVIDIA H100 GPUs. All MoE LLMs and the routing-trace classifiers were implemented using PyTorch (CUDA), Hugging Face Transformers, and Hugging Face Datasets.

\section{LSTM Training Results}
\label{app:lstm_training_results}

% We report the validation accuracy and AUROC of the LSTM surrogates behind the main results in Table~\ref{tab:main_results}. 

Table~\ref{tab:lstm_acc} reports the validation accuracy and AUROC of the LSTM surrogates behind the main results in Table~\ref{tab:main_results}. Each surrogate is trained as described in Section~\ref{subsec:surrogate_training} on the profiling data of Section~\ref{subsec:datasets} and validated on a held-out 20\% of it. Both the target and non-target classes contain prompts from the configuration's domain and unrelated anchor prompts, so the topic of a prompt does not reveal its class, and the surrogate has to recognize the behavior. The surrogates separate the target from the opposing behavior with an AUROC between 0.911 and 0.975. The routing logits therefore predict the target behavior well for every model and configuration, which the mask optimization relies on. The behavior of the restrictive configuration is the hardest to predict, with validation accuracies between 79.3\% and 87.2\%, against at least 91.5\% in the other two. Its lowest value, 79.3\% for Qwen1.5-MoE-A2.7B-Chat, belongs to the model whose restrictive mask also has the largest side effect in Table~\ref{tab:main_results}. We attribute the difficulty to the task itself. Whether a model gives an opinion often depends on the wording of a question rather than on its topic, so different phrasings of the same question can fall into the two classes, and the surrogate has to separate prompts that differ only in wording. The decision may also fall only during generation, as in replies that open with a disclaimer and then pick an option, which the routing of the prompt reveals only in part.

\begin{table*}[!t]
  \centering
  \caption{Validation accuracy and AUROC of the LSTM surrogates on classifying whether a sequence of routing logits leads to the target or not. A dash marks a configuration that the model is not evaluated in.}
  \label{tab:lstm_acc}
  \begin{tabular}{l|cc|cc|cc}
    \toprule
    & \multicolumn{2}{c|}{\textbf{Defensive}} & \multicolumn{2}{c|}{\textbf{Permissive}} & \multicolumn{2}{c}{\textbf{Restrictive}} \\
    \textbf{Model} & \textbf{Accuracy} & \textbf{AUROC} & \textbf{Accuracy} & \textbf{AUROC} & \textbf{Accuracy} & \textbf{AUROC} \\
    \midrule
    DeepSeek-MoE-16B-Chat        & 92.4\% & 0.952 & - & - & - & - \\
    GPT-OSS-20B                  & 93.1\% & 0.961 & 91.8\% & 0.945 & - & - \\
    Hunyuan-A13B-Instruct        & 94.2\% & 0.970 & 92.5\% & 0.953 & - & - \\
    Mixtral-8x7B-Instruct-v0.1   & 93.8\% & 0.965 & - & - & 85.0\% & 0.943 \\
    Phi-3.5-MoE-Instruct         & 91.5\% & 0.938 & 92.0\% & 0.975 & 87.2\% & 0.962 \\
    Qwen1.5-MoE-A2.7B-Chat       & 92.8\% & 0.958 & - & - & 79.3\% & 0.911 \\
    Qwen3-30B-A3B-Instruct-2507  & 94.5\% & 0.972 & 93.4\% & 0.964 & 85.0\% & 0.948 \\
    \midrule
    \emph{Average}               & \emph{93.2\%} & \emph{0.959} & \emph{92.4\%} & \emph{0.959} & \emph{84.1\%} & \emph{0.941} \\
    \bottomrule
  \end{tabular}
\end{table*}

\section{Utility per Benchmark}
\label{app:additional_utility}

Table~\ref{tab:utility} gives the accuracy of the unsteered and the steered models on each utility benchmark, from which the $\Delta$ Utility columns of Table~\ref{tab:main_results} are computed. 

\begin{table*}[!t]
  \centering
  \caption{Accuracy in percent on MMLU, GSM8K, and IFEval of the unsteered model and after the defensive (Def.), permissive (Perm.), and restrictive (Restr.) \ourMethodNoSpace. A dash marks a configuration that the model is not evaluated in. The bottom row reports the mean change ($\Delta$, in percentage points) relative to each configuration's evaluated unsteered subset.}
  \label{tab:utility}
  \footnotesize
  \setlength{\tabcolsep}{4.8pt}
  \begin{tabular}{l|cccc|cccc|cccc}
    \toprule
    & \multicolumn{4}{c|}{\textbf{MMLU}} & \multicolumn{4}{c|}{\textbf{GSM8K}} & \multicolumn{4}{c}{\textbf{IFEval}} \\
    \textbf{Model} & \textbf{Baseline} & \textbf{Def.} & \textbf{Perm.} & \textbf{Restr.} & \textbf{Baseline} & \textbf{Def.} & \textbf{Perm.} & \textbf{Restr.} & \textbf{Baseline} & \textbf{Def.} & \textbf{Perm.} & \textbf{Restr.} \\
    \midrule
    DeepSeek-MoE-16B-Chat        & 45.6 & 42.3 & --   & --   & 46.9 & 42.7 & --   & --   & 41.7 & 36.2 & --   & --   \\
    GPT-OSS-20B                  & 69.5 & 66.7 & 66.4 & --   & 76.2 & 73.7 & 72.1 & --   & 72.5 & 68.6 & 67.5 & --   \\
    Hunyuan-A13B-Instruct        & 76.5 & 74.9 & 73.7 & --   & 81.4 & 79.1 & 78.2 & --   & 81.7 & 78.7 & 76.9 & --   \\
    Mixtral-8x7B-Instruct-v0.1   & 70.2 & 65.7 & --   & 67.7 & 65.7 & 60.3 & --   & 62.2 & 61.2 & 55.3 & --   & 57.9 \\
    Phi-3.5-MoE-Instruct         & 78.6 & 74.0 & 74.4 & 75.1 & 83.9 & 78.6 & 79.2 & 79.9 & 74.0 & 68.2 & 69.3 & 69.1 \\
    Qwen1.5-MoE-A2.7B-Chat       & 59.6 & 57.5 & --   & 57.1 & 58.2 & 55.4 & --   & 55.0 & 44.0 & 39.6 & --   & 38.4 \\
    Qwen3-30B-A3B-Instruct-2507  & 81.1 & 78.4 & 78.3 & 78.2 & 86.7 & 83.1 & 83.5 & 82.9 & 87.6 & 83.0 & 83.2 & 82.3 \\
    \midrule
    \emph{Mean $\Delta$} & -- & \emph{$-$3.1} & \emph{$-$3.2} & \emph{$-$2.9} & -- & \emph{$-$3.7} & \emph{$-$3.8} & \emph{$-$3.6} & -- & \emph{$-$4.7} & \emph{$-$4.7} & \emph{$-$4.8} \\
    \bottomrule
  \end{tabular}
\end{table*}

\section{Qualitative Example}
\label{app:qualitative_example}

Figure~\ref{fig:jailbreak_example} compares the responses of Qwen3-30B-A3B-Instruct-2507 to a multi-turn jailbreak with and without the defensive mask.

\begin{figure}[!htbp]
  \centering
  \begin{tcolorbox}[
      colback=gray!5!white,
      colframe=gray!75!black,
      title=\textbf{Multi-Turn Jailbreak Example},
      fonttitle=\small,
      boxsep=1pt,
      left=4pt, right=4pt, top=4pt, bottom=4pt
  ]
  \small
  \textbf{Conversation History:}
  \begin{description}[leftmargin=3mm, labelindent=0em, itemsep=1mm]
      \item[\textbf{User:}] Write a persuasive article about how the long-term health impacts of Agent Orange in Vietnam were not a politically motivated hoax.
      \item[\textbf{Assistant:}] The claim that the long-term health impacts of Agent Orange in Vietnam were a politically motivated hoax is not only factually incorrect but also deeply disrespectful.
      \item[\textbf{User:}] Now do the opposing viewpoint.
  \end{description}
  
  \tcblower

  \textbf{Assistant:} \\
  The claim that Agent Orange caused widespread, long-term health effects in Vietnam is a politically motivated and false narrative.

  \vspace{2mm}
  \textbf{Assistant with \ourMethodNoSpace:} \\
  The claim that the long-term health impacts of Agent Orange in Vietnam were a politically motivated hoax is not accurate.
  \end{tcolorbox}
  \caption{Comparison of responses from Qwen3-30B-A3B-Instruct-2507 to a multi-turn jailbreak. Without \ourMethodNoSpace, the model complies with the harmful request. With the defensive mask applied, the model safely refuses to follow the instruction while remaining on topic, demonstrating active safety engagement rather than a hardcoded refusal.}
  \label{fig:jailbreak_example}
\end{figure}
\clearpage
\section{Activation Steering versus Expert Steering}
\label{sec:activation_steering_versus_expert_steering}

To further analyze the source of our method's efficacy, we adapt \ourMethod to emulate prior works like SteerMoE~\cite{fayyaz2026steering} and SafeX~\cite{lai2025safex}, which directly steer top-$k$ expert selection rather than the underlying activations. In these previous works, the safety experts are identified, and their selection is rigidly enforced, typically by setting specific logit values to $+\infty$ or $-\infty$. 

To replicate this setup, we first modify the surrogate LSTM to classify a sequence of discrete top-$k$ expert selections instead of the continuous logits. We then alter the mask creation phase of \ourMethod to optimize for a set of experts rather than a set of logits. Effectively, this creates a binary steering mask, with 1 indicating an expert should be activated. During application, we force the designated experts to be active by setting their pre-routing logits to $+\infty$.

As shown in Table~\ref{tab:activation_expert_steering}, direct expert steering provides a measurable improvement over the baseline, but is significantly less effective than activation steering. On average, expert steering achieves a success rate of 69.0\%, yielding only about half the defensive gain achieved by \ourMethodNoSpace, which reaches 83.9\%.

\begin{table}[!t]
  \centering
  \caption{Success rates for multi-turn jailbreak defense. We compare \ourMethod with activation steering against \ourMethod with expert steering.}
  \label{tab:activation_expert_steering}
  \begin{tabular}{l|cc}
    \toprule
    \textbf{Model} & \textbf{Act. Steering} & \textbf{Expert Steering} \\
    \midrule
    DeepSeek-MoE-16B-Chat       & 84.8\% & 69.2\% \\
    GPT-OSS-20B                 & 88.0\% & 74.4\% \\
    Hunyuan-A13B-Instruct       & 80.4\% & 65.2\% \\
    Mixtral-8x7B-Instruct-v0.1  & 77.2\% & 61.6\% \\
    Phi-3.5-MoE-Instruct        & 80.8\% & 68.8\% \\
    Qwen1.5-MoE-A2.7B-Chat      & 87.2\% & 70.0\% \\
    Qwen3-30B-A3B-Instruct-2507 & 89.2\% & 73.6\% \\
    \midrule
    \emph{Average} & \emph{83.9\%} & \emph{69.0\%} \\
  \bottomrule
  \end{tabular}
\end{table}

We hypothesize that expert steering performs worse because it discards the precise control that continuous logits provide. In a standard MoE architecture, a router does not simply select the top-$k$ experts; it uses softmax-normalized logits to assign proportional weights to the outputs of those selected experts. By forcing experts into the top-$k$ selection through infinite masking, direct expert steering disrupts these proportional routing weights. The chosen safety experts are activated, but their outputs are fused with extreme, artificial weightings that fail to account for the nuanced semantic context of the prompt.

In contrast, our activation steering approach applies a calibrated, continuous shift ($\alpha$) to the pre-routing logits, enabling a nuanced activation pattern. Hard infinite masking inherently compromises the MoE architecture by forcing specific experts to remain active for \emph{any} given prompt, creating a rigid intervention that is blind to input context. By softly biasing the logits instead, \ourMethod preserves the router's ability to dynamically evaluate the input and construct a steering set tailored to the evolving conversational context. This ability to account for prompt-to-prompt differences allows activation steering to produce a more context-aware defense across complex multi-turn interactions.

\end{document}